------------------------------------------------------------------------
\documentstyle[epsf]{article}
\catcode`\@=11
   \def\tableline{\hbox to \hsize}
\newbox\hdbox%
\newcount\hdrows%
\newcount\multispancount%
\newcount\ncase%
\newcount\ncols
\newcount\nrows%
\newcount\nspan%
\newcount\ntemp%
\newdimen\hdsize%
\newdimen\newhdsize%
\newdimen\parasize%
\newdimen\spreadwidth%
\newdimen\thicksize%
\newdimen\thinsize%
\newdimen\tablewidth%
\newif\ifcentertables%
\newif\ifendsize%
\newif\iffirstrow%
\newif\iftableinfo%
\newtoks\dbt%
\newtoks\hdtks%
\newtoks\savetks%
\newtoks\tableLETtokens%
\newtoks\tabletokens%
\newtoks\widthspec%
\tableinfotrue%
\def\tstrut{\vrule height3.1ex depth1.2ex width0pt}%
\def\and{\char`\&}
\def\tablerule{\noalign{\hrule height\thinsize depth0pt}}%
\def\thickrule{\noalign{\hrule height\thicksize depth0pt}}%
\def\ctr#1{\hfil\ #1\hfil}%
\tablewidth=-\maxdimen%
\spreadwidth=-\maxdimen%
\def\tabskipglue{0pt plus 1fil minus 1fil}%
\centertablestrue%
\gdef\ARGS{########}
\gdef\headerARGS{####}
\def\@mpersand{&}
{\catcode`\|=13
\gdef\letbarzero{\let|0}
\gdef\letbartab{\def|{&&}}%
\gdef\letvbbar{\let\vb|}%
}
{\catcode`\&=4
\def\ampskip{&\omit\hfil&}
\catcode`\&=13
\let&0
\xdef\letampskip{\def&{\ampskip}}%
\gdef\letnovbamp{\let\novb&\let\tab&}
}
\def\begintable{
   \begingroup%
   \catcode`\|=13\letbartab\letvbbar%
   \catcode`\&=13\letampskip\letnovbamp%
   \def\multispan##1{
      \omit \mscount##1%
      \multiply\mscount\tw@\advance\mscount\m@ne%
      \loop\ifnum\mscount>\@ne \sp@n\repeat%
   }
   \def\|{%
      &\omit\widevline&%
   }%
   \ruledtable
}
\long\def\ruledtable#1\endtable{%
   \offinterlineskip
   \tabskip 0pt
   \def\widevline{\vrule width\thicksize}
   \def\endrow{\@mpersand\omit\hfil\crnorm\@mpersand}%
   \def\crthick{\@mpersand\crnorm\thickrule\@mpersand}%
   \def\crnorule{\@mpersand\crnorm\@mpersand}%
   \let\nr=\crnorule
   \def\endtable{\@mpersand\crnorm\thickrule}%
   \let\crnorm=\cr
   \edef\cr{\@mpersand\crnorm\tablerule\@mpersand}%
   \the\tableLETtokens
   \tabletokens={&#1}
   \countROWS\tabletokens\into\nrows%
   \countCOLS\tabletokens\into\ncols%
   \advance\ncols by -1%
   \divide\ncols by 2%
   \advance\nrows by 1%
   \iftableinfo %
      \immediate\write16{[Nrows=\the\nrows, Ncols=\the\ncols]}%
   \fi%
   \ifcentertables
      \ifhmode \par\fi
      \tableline{
      \hss
   \else %
      \hbox{%
   \fi
      \vbox{%
         \makePREAMBLE{\the\ncols}
         \edef\next{\preamble}
         \let\preamble=\next
         \makeTABLE{\preamble}{\tabletokens}
      }
      \ifcentertables \hss}\else }\fi
   \endgroup
   \tablewidth=-\maxdimen
   \spreadwidth=-\maxdimen
}
\def\makeTABLE#1#2{
   {
   \let\ifmath0
   \let\header0
   \let\multispan0
   \ncase=0%
   \ifdim\tablewidth>-\maxdimen \ncase=1\fi%
   \ifdim\spreadwidth>-\maxdimen \ncase=2\fi%
   \relax
   \ifcase\ncase %
      \widthspec={}%
   \or %
      \widthspec=\expandafter{\expandafter t\expandafter o%
                 \the\tablewidth}%
   \else %
      \widthspec=\expandafter{\expandafter s\expandafter p\expandafter r%
                 \expandafter e\expandafter a\expandafter d%
                 \the\spreadwidth}%
   \fi %
   \xdef\next{
      \halign\the\widthspec{%
      #1
      \noalign{\hrule height\thicksize depth0pt}
      \the#2\endtable
      }
   }
   }
   \next
}
\def\makePREAMBLE#1{
   \ncols=#1
   \begingroup
   \let\ARGS=0
   \edef\xtp{\widevline\ARGS\tabskip\tabskipglue%
   &\ctr{\ARGS}\tstrut}
   \advance\ncols by -1
   \loop
      \ifnum\ncols>0 %
      \advance\ncols by -1%
      \edef\xtp{\xtp&\vrule width\thinsize\ARGS&\ctr{\ARGS}}%
   \repeat
   \xdef\preamble{\xtp&\widevline\ARGS\tabskip0pt%
   \crnorm}
   \endgroup
}
\def\countROWS#1\into#2{
   \let\countREGISTER=#2%
   \countREGISTER=0%
   \expandafter\ROWcount\the#1\endcount%
}%
\def\ROWcount{%
   \afterassignment\subROWcount\let\next= %
}%
\def\subROWcount{%
   \ifx\next\endcount %
      \let\next=\relax%
   \else%
      \ncase=0%
      \ifx\next\cr %
         \global\advance\countREGISTER by 1%
         \ncase=0%
      \fi%
      \ifx\next\endrow %
         \global\advance\countREGISTER by 1%
         \ncase=0%
      \fi%
      \ifx\next\crthick %
         \global\advance\countREGISTER by 1%
         \ncase=0%
      \fi%
      \ifx\next\crnorule %
         \global\advance\countREGISTER by 1%
         \ncase=0%
      \fi%
      \ifx\next\header %
         \ncase=1%
      \fi%
      \relax%
      \ifcase\ncase %
         \let\next\ROWcount%
      \or %
         \let\next\argROWskip%
      \else %
      \fi%
   \fi%
   \next%
}
\def\counthdROWS#1\into#2{%
\dvr{10}%
   \let\countREGISTER=#2%
   \countREGISTER=0%
\dvr{11}%
\dvr{13}%
   \expandafter\hdROWcount\the#1\endcount%
\dvr{12}%
}%
\def\hdROWcount{%
   \afterassignment\subhdROWcount\let\next= %
}%
\def\subhdROWcount{%
   \ifx\next\endcount %
      \let\next=\relax%
   \else%
      \ncase=0%
      \ifx\next\cr %
         \global\advance\countREGISTER by 1%
         \ncase=0%
      \fi%
      \ifx\next\endrow %
         \global\advance\countREGISTER by 1%
         \ncase=0%
      \fi%
      \ifx\next\crthick %
         \global\advance\countREGISTER by 1%
         \ncase=0%
      \fi%
      \ifx\next\crnorule %
         \global\advance\countREGISTER by 1%
         \ncase=0%
      \fi%
      \ifx\next\header %
         \ncase=1%
      \fi%
\relax%
      \ifcase\ncase %
         \let\next\hdROWcount%
      \or%
         \let\next\arghdROWskip%
      \else %
      \fi%
   \fi%
   \next%
}%
{\catcode`\|=13\letbartab
\gdef\countCOLS#1\into#2{%
   \let\countREGISTER=#2%
   \global\countREGISTER=0%
   \global\multispancount=0%
   \global\firstrowtrue
   \expandafter\COLcount\the#1\endcount%
   \global\advance\countREGISTER by 3%
   \global\advance\countREGISTER by -\multispancount
}%
\gdef\COLcount{%
   \afterassignment\subCOLcount\let\next= %
}%
{\catcode`\&=13%
\gdef\subCOLcount{%
   \ifx\next\endcount %
      \let\next=\relax%
   \else%
      \ncase=0%
      \iffirstrow
         \ifx\next& %
            \global\advance\countREGISTER by 2%
            \ncase=0%
         \fi%
         \ifx\next\span %
            \global\advance\countREGISTER by 1%
            \ncase=0%
         \fi%
         \ifx\next| %
            \global\advance\countREGISTER by 2%
            \ncase=0%
         \fi
         \ifx\next\|
            \global\advance\countREGISTER by 2%
            \ncase=0%
         \fi
         \ifx\next\multispan
            \ncase=1%
            \global\advance\multispancount by 1%
         \fi
         \ifx\next\header
            \ncase=2%
         \fi
         \ifx\next\cr       \global\firstrowfalse \fi
         \ifx\next\endrow   \global\firstrowfalse \fi
         \ifx\next\crthick  \global\firstrowfalse \fi
         \ifx\next\crnorule \global\firstrowfalse \fi
      \fi
\relax
      \ifcase\ncase %
         \let\next\COLcount%
      \or %
         \let\next\spancount%
      \or %
         \let\next\argCOLskip%
      \else %
      \fi %
   \fi%
   \next%
}%
\gdef\argROWskip#1{%
   \let\next\ROWcount \next%
}
\gdef\arghdROWskip#1{%
   \let\next\ROWcount \next%
}
\gdef\argCOLskip#1{%
   \let\next\COLcount \next%
}
}
}
\def\spancount#1{
   \nspan=#1\multiply\nspan by 2\advance\nspan by -1%
   \global\advance \countREGISTER by \nspan
   \let\next\COLcount \next}%
\def\dvr#1{\relax}%
\def\header#1{%
\dvr{1}{\let\cr=\@mpersand%
\hdtks={#1}%
\counthdROWS\hdtks\into\hdrows%
\advance\hdrows by 1%
\ifnum\hdrows=0 \hdrows=1 \fi%
\dvr{5}\makehdPREAMBLE{\the\hdrows}%
\dvr{6}\getHDdimen{#1}%
{\parindent=0pt\hsize=\hdsize{\let\ifmath0%
\xdef\next{\valign{\headerpreamble #1\crnorm}}}\dvr{7}\next\dvr{8}%
}%
}\dvr{2}}
\def\makehdPREAMBLE#1{
\dvr{3}%
\hdrows=#1
{
\let\headerARGS=0%
\let\cr=\crnorm%
\edef\xtp{\vfil\hfil\hbox{\headerARGS}\hfil\vfil}%
\advance\hdrows by -1
\loop
\ifnum\hdrows>0%
\advance\hdrows by -1%
\edef\xtp{\xtp&\vfil\hfil\hbox{\headerARGS}\hfil\vfil}%
\repeat%
\xdef\headerpreamble{\xtp\crcr}%
}
\dvr{4}}
\def\getHDdimen#1{%
\hdsize=0pt%
\getsize#1\cr\end\cr%
}
\def\getsize#1\cr{%
\endsizefalse\savetks={#1}%
\expandafter\lookend\the\savetks\cr%
\relax \ifendsize \let\next\relax \else%
\setbox\hdbox=\hbox{#1}\newhdsize=1.0\wd\hdbox%
\ifdim\newhdsize>\hdsize \hdsize=\newhdsize \fi%
\let\next\getsize \fi%
\next%
}%
\def\lookend{\afterassignment\sublookend\let\looknext= }%
\def\sublookend{\relax%
\ifx\looknext\cr %
\let\looknext\relax \else %
   \relax
   \ifx\looknext\end \global\endsizetrue \fi%
   \let\looknext=\lookend%
    \fi \looknext%
}%
\def\tablelet#1{%
   \tableLETtokens=\expandafter{\the\tableLETtokens #1}%
}%
\newdimen\Parindent\newdimen\Parskip
\Parindent=\parindent\Parskip=\parskip
\def\@oddhead{}\def\@evenhead{}
\def\ps@myheadings{\let\@mkboth\@gobbletwo
\def\@oddhead{\hbox{}\sl\rightmark}
\def\@oddfoot{\hfil}\def\@evenhead{\hfil\sl\leftmark\hbox
{}}\def\@evenfoot{\hfil}\def\sectionmark##1{}\def\subsectionmark##1{}}

\newdimen\Parindent\newdimen\Parskip
{\Parindent=\parindent\Parskip=\parskip
\parindent0pt\parskip3mm\columnsep11mm}
{\parindent\Parindent\parskip\Parskip}
{\Parindent=\parindent\Parskip=\parskip
\parindent0pt\parskip3mm\columnsep11mm}
{\parindent\Parindent\parskip\Parskip}

\gdef\abstract#1{\gdef\@abstract{#1}}

\def\maketitle{\par
 \begingroup
 \def\thefootnote{\fnsymbol{footnote}}
 \def\@makefnmark{\hbox
 to 0pt{$^{\@thefnmark}$\hss}}
 \twocolumn[\@maketitle]
 \@thanks
 \endgroup
 \setcounter{footnote}{0}
 \let\maketitle\relax
 \let\@maketitle\relax
 \gdef\@thanks{}\gdef\@author{}\gdef\@title{}\let\thanks\relax}
\def\@maketitle{\vbox to 1in{\hsize\textwidth
 \linewidth\hsize \vfil \centering
 { \Large\bf \@title \par} \vskip 1.3em {\begin{tabular}[t]{c}\@author
\end{tabular}\par}
 \vfil}
\hsize\textwidth \linewidth\hsize
\begin{center} ABSTRACT \end{center} \par \small\rm \@abstract \vskip 2em }

\def\section{\@startsection {section}{1}{\z@}{-3.5ex plus -1ex minus
 -.2ex}{2.3ex plus .2ex}{\normalsize\bf}}
\def\subsection{\@startsection{subsection}{2}{\z@}{-3.25ex plus -1ex minus
 -.2ex}{1.5ex plus .2ex}{\normalsize\bf}}
\def\subsubsection{\@startsection{subsubsection}{3}{\z@}{-3.25ex plus
-1ex minus -.2ex}{1.5ex plus .2ex}{\normalsize\bf}}
\def\paragraph{\@startsection
 {paragraph}{4}{\z@}{3.25ex plus 1ex minus .2ex}{-1em}{\normalsize\bf}}
\def\subparagraph{\@startsection
 {subparagraph}{4}{\parindent}{3.25ex plus 1ex minus
 .2ex}{-1em}{\normalsize\bf}}
\catcode`\@=12
\def\sto{\!\to\!}
\def\diha{$\Delta I\sheq 1/2$}
\def\sheq{\!=\!}
\def\particleone{\rm}
\def\particletwo{\sl}

\def\tiny{\vrule width 0pt}
\def\star{{\bf *}}
\def\conventionone{

 \def\PM{\relax\ifmmode{\pm}\else{$\pm$}\fi}

 \def\decays{\relax\ifmmode{\rightarrow}\else{$\rightarrow$}\fi\tiny}

 \def\EPEM{\relax\ifmmode{e^+e^-}\else{$e^+e^-$}\fi}
 \def\epem{\relax\ifmmode{e^+e^-}\else{$e^+e^-$}\fi}

 \def\G{\relax\ifmmode{\gamma}\else{$\gamma$}\fi}

 \def\W{\relax\ifmmode{{\particleone W}}\else{{\particleone W}}\fi}
 \def\WP{\relax\ifmmode{{\particleone W}^+}\else{{\particleone W}$^+$}\fi}
 \def\WM{\relax\ifmmode{{\particleone W}^-}\else{{\particleone W}$^-$}\fi}
 \def\WPM{\relax\ifmmode{{\particleone W}^\pm}\else{{\particleone W}$^\pm$}\fi}
 \def\WMP{\relax\ifmmode{{\particleone W}^\mp}\else{{\particleone W}$^\mp$}\fi}

 \def\Z{\relax\ifmmode{{\particleone Z}}\else{{\particleone Z}}\fi}
 \def\ZZ{\relax\ifmmode{{\particleone Z}^0}\else{{\particleone Z}$^0$}\fi}

 \def\NU{\relax\ifmmode{\nu}\else{$\nu$}\fi}
 \def\NUB{\relax\ifmmode{\overline{\nu}}
	\else{$\overline{\nu}$}\fi}

 \def\NE{\relax\ifmmode{\nu_e}\else{$\nu_e$}\fi}
 \def\NEB{\relax\ifmmode{\overline{\nu}\tiny_e}
	\else{$\overline{\nu}\tiny_e$}\fi}

 \def\E{\relax\ifmmode{e}\else{$e$}\fi}
 \def\EP{\relax\ifmmode{e^+}\else{$e^+$}\fi}
 \def\EM{\relax\ifmmode{e^-}\else{$e^-$}\fi}
 \def\EPM{\relax\ifmmode{e^\pm}\else{$e^\pm$}\fi}
 \def\EMP{\relax\ifmmode{e^\mp}\else{$e^\mp$}\fi}

 \def\NM{\relax\ifmmode{\nu_\mu}\else{$\nu_\mu$}\fi}
 \def\NMB{\relax\ifmmode{\overline{\nu}\tiny_\mu}
	\else{$\overline{\nu}\tiny_\mu$}\fi}

 \def\M{\relax\ifmmode{\mu}\else{$\mu$}\fi}
 \def\MP{\relax\ifmmode{\mu^+}\else{$\mu^+$}\fi}
 \def\MM{\relax\ifmmode{\mu^-}\else{$\mu^-$}\fi}
 \def\MPM{\relax\ifmmode{\mu^\pm}\else{$\mu^\pm$}\fi}
 \def\MMP{\relax\ifmmode{\mu^\mp}\else{$\mu^\mp$}\fi}

 \def\NT{\relax\ifmmode{\nu_\tau}\else{$\nu_\tau$}\fi}
 \def\NTB{\relax\ifmmode{\overline{\nu}\tiny_\tau}
	\else{$\overline{\nu}\tiny_\tau$}\fi}

 \def\T{\relax\ifmmode{\tau}\else{$\tau$}\fi}
 \def\TP{\relax\ifmmode{\tau^+}\else{$\tau^+$}\fi}
 \def\TM{\relax\ifmmode{\tau^-}\else{$\tau^-$}\fi}
 \def\TPM{\relax\ifmmode{\tau^\pm}\else{$\tau^\pm$}\fi}
 \def\TMP{\relax\ifmmode{\tau^\mp}\else{$\tau^\mp$}\fi}

 \def\NL{\relax\ifmmode{\nu_\ell}\else{$\nu_\ell$}\fi}
 \def\NLB{\relax\ifmmode{\overline{\nu}\tiny_\ell}
	\else{$\overline{\nu}\tiny_\ell$}\fi}

 \def\L{\relax\ifmmode{\ell}\else{$\ell$}\fi}
 \def\LP{\relax\ifmmode{\ell^+}\else{$\ell^+$}\fi}
 \def\LM{\relax\ifmmode{\ell^-}\else{$\ell^-$}\fi}
 \def\LPM{\relax\ifmmode{\ell^\pm}\else{$\ell^\pm$}\fi}
 \def\LMP{\relax\ifmmode{\ell^\mp}\else{$\ell^\mp$}\fi}

 \def\PI{\relax\ifmmode{\pi}\else{$\pi$}\fi}
 \def\PIP{\relax\ifmmode{\pi^+}\else{$\pi^+$}\fi}
 \def\PIZ{\relax\ifmmode{\pi^0}\else{$\pi^0$}\fi}
 \def\PIM{\relax\ifmmode{\pi^-}\else{$\pi^-$}\fi}
 \def\PIPM{\relax\ifmmode{\pi^\pm}\else{$\pi^\pm$}\fi}
 \def\PIMP{\relax\ifmmode{\pi^\mp}\else{$\pi^\mp$}\fi}
 \def\PIPMZ{\relax\ifmmode{\pi^{\pm,0}}\else{$\pi^{\pm,0}$}\fi}

 \def\ET{\relax\ifmmode{\eta}\else{$\eta$}\fi}
 \def\ETZ{\relax\ifmmode{\eta^0}\else{$\eta^0$}\fi}

 \def\K{\relax\ifmmode{{\particleone K}}\else{{\particleone K}}\fi}
 \def\KB{\relax\ifmmode{\overline{{\particleone K}}}
	\else{$\overline{{\particleone K}}$}\fi}

 \def\KZ{\relax\ifmmode{{\particleone K}^0}\else{{\particleone K}$^0$}\fi}
 \def\KSH{\relax\ifmmode{{\particleone K}^0_S}\else{{\particleone K}$^0_S$}\fi}
 \def\KLO{\relax\ifmmode{{\particleone K}^0_L}\else{{\particleone K}$^0_L$}\fi}
 \def\KZB{\relax\ifmmode{\overline{{\particleone K}}\tiny^0}
	\else{$\overline{{\particleone K}}\tiny^0$}\fi}

 \def\KP{\relax\ifmmode{{\particleone K}^+}\else{{\particleone K}$^+$}\fi}
 \def\KM{\relax\ifmmode{{\particleone K}^-}\else{{\particleone K}$^-$}\fi}
 \def\KPM{\relax\ifmmode{{\particleone K}^\pm}\else{{\particleone K}$^\pm$}\fi}
 \def\KMP{\relax\ifmmode{{\particleone K}^\mp}\else{{\particleone K}$^\mp$}\fi}

 \def\D{\relax\ifmmode{{\particleone D}}\else{{\particleone D}}\fi\tiny}
 \def\DB{\relax\ifmmode{\overline{{\particleone D}}}
	\else{$\overline{{\particleone D}}$}\fi}

 \def\DZ{\relax\ifmmode{{\particleone D}^0}\else{{\particleone D}$^0$}\fi}
 \def\DZB{\relax\ifmmode{\overline{{\particleone D}}\tiny^0}
	\else{$\overline{{\particleone D}}\tiny^0$}\fi}

 \def\DP{\relax\ifmmode{{\particleone D}^+}\else{{\particleone D}$^+$}\fi}
 \def\DM{\relax\ifmmode{{\particleone D}^-}\else{{\particleone D}$^-$}\fi}
 \def\DPM{\relax\ifmmode{{\particleone D}^\pm}\else{{\particleone D}$^\pm$}\fi}
 \def\DMP{\relax\ifmmode{{\particleone D}^\mp}\else{{\particleone D}$^\mp$}\fi}

 \def\F{\relax\ifmmode{{\particleone F}}\else{{\particleone F}}\fi}
 \def\FP{\relax\ifmmode{{\particleone F}^+}\else{{\particleone F}$^+$}\fi}
 \def\FM{\relax\ifmmode{{\particleone F}^-}\else{{\particleone F}$^-$}\fi}
 \def\FPM{\relax\ifmmode{{\particleone F}^\pm}\else{{\particleone F}$^\pm$}\fi}
 \def\FMP{\relax\ifmmode{{\particleone F}^\mp}\else{{\particleone F}$^\mp$}\fi}

 \def\DUS{\relax\ifmmode{{\particletwo D}_s}\else{{\particletwo D}$_s$}\fi}
 \def\DUSO{\relax\ifmmode{{\particletwo D}_{s1}}
      \else{{\particletwo D}$_{s1}$}\fi}
 \def\DUSP{\relax\ifmmode{{\particletwo D}_s^+}
      \else{{\particletwo D}$^+$}\fi}
 \def\DUSM{\relax\ifmmode{{\particletwo D}_s^-}
      \else{{\particletwo D}$^-$}\fi}
 \def\DUSPM{\relax\ifmmode{{\particletwo D}_s^\pm}
           \else{{\particletwo D}$_s^\pm$}\fi}
 \def\DUSMP{\relax\ifmmode{{\particletwo D}_s^\mp}
           \else{{\particletwo D}$_s^\mp$}\fi}

 \def\B{\relax\ifmmode{{\particleone B}}\else{{\particleone B}}\fi\tiny}
 \def\BB{\relax\ifmmode{\overline{{\particleone B}}}
	\else{$\overline{{\particleone B}}$}\fi}

 \def\BZ{\relax\ifmmode{{\particleone B}^0}\else{{\particleone B}$^0$}\fi}
 \def\BZB{\relax\ifmmode{\overline{{\particleone B^0}}}
	\else{$\overline{{\particleone B}^0}$}\fi}

 \def\BP{\relax\ifmmode{{\particleone B}^+}\else{{\particleone B}$^+$}\fi}
 \def\BM{\relax\ifmmode{{\particleone B}^-}\else{{\particleone B}$^-$}\fi}
 \def\BPM{\relax\ifmmode{{\particleone B}^\pm}\else{{\particleone B}$^\pm$}\fi}
 \def\BMP{\relax\ifmmode{{\particleone B}^\mp}\else{{\particleone B}$^\mp$}\fi}

 \def\eM{\relax\ifmmode{{\particleone M}}\else{{\particleone M}}\fi\tiny}
 \def\eMB{\relax\ifmmode{\overline{{\particleone M}}}
	\else{$\overline{{\particleone M}}$}\fi}

 \def\eMZ{\relax\ifmmode{{\particleone M}^0}\else{{\particleone M}$^0$}\fi}
 \def\eMZB{\relax\ifmmode{\overline{{\particleone M}}\tiny^0}
	\else{$\overline{{\particleone M}}\tiny^0$}\fi}

 \def\eMP{\relax\ifmmode{{\particleone M}^+}\else{{\particleone M}$^+$}\fi}
 \def\eMM{\relax\ifmmode{{\particleone M}^-}\else{{\particleone M}$^-$}\fi}
 \def\eMPM{\relax\ifmmode{{\particleone M}^\pm}
   \else{{\particleone M}$^\pm$}\fi}
 \def\eMMP{\relax\ifmmode{{\particleone M}^\mp}
   \else{{\particleone M}$^\mp$}\fi}

 \def\PR{\relax\ifmmode{{\particleone p}}\else{{\particleone p}}\fi}
 \def\PB{\relax\ifmmode{\overline{{\particleone p}}}
	\else{$\overline{{\particleone p}}$}\fi}

 \def\NR{\relax\ifmmode{{\particleone n}}\else{{\particleone n}}\fi}
 \def\NB{\relax\ifmmode{\overline{{\particleone n}}}
	\else{$\overline{{\particleone n}}$}\fi}

 \def\LA{\relax\ifmmode{\Lambda}\else{$\Lambda$}\fi}
 \def\LAB{\relax\ifmmode{\overline{\Lambda}}
	\else{$\overline{\Lambda}$}\fi}
 \def\LAZ{\relax\ifmmode{\Lambda^0}\else{$\Lambda^0$}\fi}
 \def\LAZB{\relax\ifmmode{\overline{\Lambda}\tiny^0}
	\else{$\overline{\Lambda}\tiny^0$}\fi}

 \def\SI{\relax\ifmmode{\Sigma}\else{$\Sigma$}\fi}
 \def\SIB{\relax\ifmmode{\overline{\Sigma}}
	\else{$\overline{\Sigma}$}\fi}
 \def\SIZ{\relax\ifmmode{\Sigma^0}\else{$\Sigma^0$}\fi}
 \def\SIZB{\relax\ifmmode{\overline{\Sigma}\tiny^0}
	\else{$\overline{\Sigma}\tiny^0$}\fi}
 \def\SIP{\relax\ifmmode{\Sigma^+}\else{$\Sigma^+$}\fi}
 \def\SIM{\relax\ifmmode{\Sigma^-}\else{$\Sigma^-$}\fi}
 \def\SIPM{\relax\ifmmode{\Sigma^\pm}\else{$\Sigma^\pm$}\fi}
 \def\SIMP{\relax\ifmmode{\Sigma^\mp}\else{$\Sigma^\mp$}\fi}
 \def\SIPB{\relax\ifmmode{\overline{\Sigma}\tiny^+}
	\else{$\overline{\Sigma}\tiny^+$}\fi}
 \def\SIMB{\relax\ifmmode{\overline{\Sigma}\tiny^-}
	\else{$\overline{\Sigma}\tiny^-$}\fi}
 \def\SIPMB{\relax\ifmmode{\overline{\Sigma}\tiny^\pm}
	\else{$\overline{\Sigma}\tiny^\pm$}\fi}
 \def\SIMPB{\relax\ifmmode{\overline{\Sigma}\tiny^\mp}
	\else{$\overline{\Sigma}\tiny^\mp$}\fi}

 \def\XI{\relax\ifmmode{\Xi}\else{$\Xi$}\fi}
 \def\XIB{\relax\ifmmode{\overline{\Xi}}
	\else{$\overline{\Xi}$}\fi}
 \def\XIZ{\relax\ifmmode{\Xi^0}\else{$\Xi^0$}\fi}
 \def\XIZB{\relax\ifmmode{\overline{\Xi}\tiny^0}
	\else{$\overline{\Xi}\tiny^0$}\fi}
 \def\XIM{\relax\ifmmode{\Xi^-}\else{$\Xi^-$}\fi}
 \def\XIPB{\relax\ifmmode{\overline{\Xi}\tiny^+}
	\else{$\overline{\Xi}\tiny^+$}\fi}

 \def\OMM{\relax\ifmmode{\Omega^-}\else{$\Omega^-$}\fi}
 \def\OMPB{\relax\ifmmode{\overline{\Omega}\tiny^+}
	\else{$\overline{\Omega}\tiny^+$}\fi}

 \def\LC{\relax\ifmmode{\Lambda_c}\else{$\Lambda_c$}\fi}
 \def\LCB{\relax\ifmmode{\overline{\Lambda_c}}
	\else{$\overline{\Lambda}\tiny_c$}\fi}
 \def\LCP{\relax\ifmmode{\Lambda^+_c}\else{$\Lambda^+_c$}\fi}
 \def\LCMB{\relax\ifmmode{\overline{\Lambda}\tiny^-_c}
	\else{$\overline{\Lambda}\tiny^-_c$}\fi}

 \def\RH{\relax\ifmmode{\rho}\else{$\rho$}\fi}
 \def\RHP{\relax\ifmmode{\rho^+}\else{$\rho^+$}\fi}
 \def\RHZ{\relax\ifmmode{\rho^0}\else{$\rho^0$}\fi}
 \def\RHM{\relax\ifmmode{\rho^-}\else{$\rho^-$}\fi}
 \def\RHPM{\relax\ifmmode{\rho^\pm}\else{$\rho^\pm$}\fi}
 \def\RHMP{\relax\ifmmode{\rho^\mp}\else{$\rho^\mp$}\fi}
 \def\RHPMZ{\relax\ifmmode{\rho^{\pm,0}}\else{$\rho^{\pm,0}$}\fi}

 \def\oM{\relax\ifmmode{\omega}\else{$\omega$}\fi}
 \def\oMZ{\relax\ifmmode{\omega^0}\else{$\omega^0$}\fi}

 \def\ETP{\relax\ifmmode{\eta'}\else{$\eta'$}\fi}
 \def\ETPZ{\relax\ifmmode{\eta'\tiny^0}\else{$\eta'\tiny^0$}\fi}

 \def\PH{\relax\ifmmode{\phi}\else{$\phi$}\fi}

 \def\PS{\relax\ifmmode{J/\psi}\else{$J/\psi$}\fi}
 \def\PSP{\relax\ifmmode{\psi'}\else{$\psi'$}\fi}
 \def\PSPP{\relax\ifmmode{\psi''}\else{$\psi''$}\fi}
 \def\PSPPP{\relax\ifmmode{\psi'''}\else{$\psi'''$}\fi}
 \def\PSPPPP{\relax\ifmmode{\psi''''}\else{$\psi''''$}\fi}

 \def\US{\relax\ifmmode{\Upsilon{\particleone (1S)}}
	\else{$\Upsilon{\particleone (1S)}$}\fi}
 \def\USS{\relax\ifmmode{\Upsilon{\particleone (2S)}}
	\else{$\Upsilon{\particleone (2S)}$}\fi}
 \def\USSS{\relax\ifmmode{\Upsilon{\particleone (3S)}}
	\else{$\Upsilon{\particleone (3S)}$}\fi}
 \def\USSSS{\relax\ifmmode{\Upsilon{\particleone (4S)}}
	\else{$\Upsilon{\particleone (4S)}$}\fi}
 \def\USSSSS{\relax\ifmmode{\Upsilon{\particleone (5S)}}
	\else{$\Upsilon{\particleone (5S)}$}\fi}
 \def\USSSSSS{\relax\ifmmode{\Upsilon{\particleone (6S)}}
	\else{$\Upsilon{\particleone (6S)}$}\fi}

 \def\KS{\relax\ifmmode{{\particleone K}^\star}
  \else{{\particleone K}$^\star$}\fi}
 \def\KSB{\relax\ifmmode{\overline{{\particleone K}}\tiny^\star}
	\else{$\overline{{\particleone K}}\tiny^\star$}\fi}

 \def\KSZ{\relax\ifmmode{{\particleone K}^{\star0}}
  \else{{\particleone K}$^{\star0}$}\fi}
 \def\KSZB{\relax\ifmmode{\overline{{\particleone K}}\tiny^{\star0}}
	\else{$\overline{{\particleone K}}\tiny^{\star0}$}\fi}

 \def\KSP{\relax\ifmmode{{\particleone K}^{\star+}}
  \else{{\particleone K}$^{\star+}$}\fi}
 \def\KSM{\relax\ifmmode{{\particleone K}^{\star-}}
  \else{{\particleone K}$^{\star-}$}\fi}
 \def\KSPM{\relax\ifmmode{{\particleone K}^{\star\pm}}
  \else{{\particleone K}$^{\star\pm}$}\fi}
 \def\KSMP{\relax\ifmmode{{\particleone K}^{\star\mp}}
\else{{\particleone K}$^{\star\mp}$}\fi}

 \def\DS{\relax\ifmmode{{\particleone D}^\star}
\else{{\particleone D}$^\star$}\fi}
 \def\DSB{\relax\ifmmode{\overline{{\particleone D}}\tiny^\star}
	\else{$\overline{{\particleone D}}\tiny^\star$}\fi}

 \def\DSZ{\relax\ifmmode{{\particleone D}^{\star0}}
\else{{\particleone D}$^{\star0}$}\fi}
 \def\DSZB{\relax\ifmmode{\overline{{\particleone D}}\tiny^{\star0}}
	\else{$\overline{{\particleone D}}\tiny^{\star0}$}\fi}

 \def\DSP{\relax\ifmmode{{\particleone D}^{\star+}}
\else{{\particleone D}$^{\star+}$}\fi}
 \def\DSM{\relax\ifmmode{{\particleone D}^{\star-}}
\else{{\particleone D}$^{\star-}$}\fi}
 \def\DSPM{\relax\ifmmode{{\particleone D}^{\star\pm}}
\else{{\particleone D}$^{\star\pm}$}\fi}
 \def\DSMP{\relax\ifmmode{{\particleone D}^{\star\mp}}
\else{{\particleone D}$^{\star\mp}$}\fi}

 \def\DDS{\relax\ifmmode{{\particleone D}^{\star\star}}
\else{{\particleone D}$^{\star\star}$}\fi}
 \def\DDSB{\relax\ifmmode{\overline{{\particleone D}}\tiny^{\star\star}}
	\else{$\overline{{\particleone D}}\tiny^{\star\star}$}\fi}

 \def\DDSZ{\relax\ifmmode{{\particleone D}^{\star\star0}}
\else{{\particleone D}$^{\star\star0}$}\fi}
 \def\DDSZB{\relax\ifmmode{\overline{{\particleone D}}\tiny^{\star\star0}}
	\else{$\overline{{\particleone D}}\tiny^{\star\star0}$}\fi}

 \def\DDSP{\relax\ifmmode{{\particleone D}^{\star\star+}}
\else{{\particleone D}$^{\star\star+}$}\fi}
 \def\DDSM{\relax\ifmmode{{\particleone D}^{\star\star-}}
\else{{\particleone D}$^{\star\star-}$}\fi}
 \def\DDSPM{\relax\ifmmode{{\particleone D}^{\star\star\pm}}
	\else{{\particleone D}$^{\star\star\pm}$}\fi}
 \def\DDSMP{\relax\ifmmode{{\particleone D}^{\star\star\mp}}
	\else{{\particleone D}$^{\star\star\mp}$}\fi}

}

\def\conventiontwo{

 \def\epem{\relax\ifmmode{e^+e^-}\else{$e^+e^-$}\fi}

 \def\decays{\relax\ifmmode{\rightarrow}\else{$\rightarrow$}\fi\tiny}

\def\gamma{\relax\ifmmode{\mathchar"10D}\else{$\mathchar"10D$}\fi}

 \def\W{\relax\ifmmode{{\particletwo W}}\else{{\particletwo W}}\fi}
 \def\Wplus{\relax\ifmmode{{\particletwo W}^+}\else{{\particletwo W}$^+$}\fi}
 \def\Wminus{\relax\ifmmode{{\particletwo W}^-}\else{{\particletwo W}$^-$}\fi}
 \def\Wpm{\relax\ifmmode{{\particletwo W}^\pm}\else{{\particletwo W}$^\pm$}\fi}
 \def\Wmp{\relax\ifmmode{{\particletwo W}^\mp}\else{{\particletwo W}$^\mp$}\fi}

 \def\Z{\relax\ifmmode{{\particletwo Z}}\else{{\particletwo Z}}\fi}
 \def\Zzero{\relax\ifmmode{{\particletwo Z}^0}\else{{\particletwo Z}$^0$}\fi}

 \def\nu{\relax\ifmmode{\mathchar"117}\else{$\mathchar"117$}\fi}
 \def\nubar{\relax\ifmmode{\overline{\nu}}
	\else{$\overline{\nu}$}\fi}

 \def\nue{\relax\ifmmode{\nu_e}\else{$\nu_e$}\fi}
 \def\nuebar{\relax\ifmmode{\overline{\nu}\tiny_e}
	\else{$\overline{\nu}\tiny_e$}\fi}

 \def\e{\relax\ifmmode{e}\else{$e$}\fi}
 \def\eplus{\relax\ifmmode{e^+}\else{$e^+$}\fi}
 \def\eminus{\relax\ifmmode{e^-}\else{$e^-$}\fi}
 \def\epm{\relax\ifmmode{e^\pm}\else{$e^\pm$}\fi}
 \def\emp{\relax\ifmmode{e^\mp}\else{$e^\mp$}\fi}

 \def\numu{\relax\ifmmode{\nu_\mu}\else{$\nu_\mu$}\fi}
 \def\numubar{\relax\ifmmode{\overline{\nu}\tiny_\mu}
	\else{$\overline{\nu}\tiny_\mu$}\fi}

 \def\mu{\relax\ifmmode{\mathchar"116}\else{$\mathchar"116$}\fi}
 \def\muplus{\relax\ifmmode{\mu^+}\else{$\mu^+$}\fi}
 \def\muminus{\relax\ifmmode{\mu^-}\else{$\mu^-$}\fi}
 \def\mupm{\relax\ifmmode{\mu^\pm}\else{$\mu^\pm$}\fi}
 \def\mump{\relax\ifmmode{\mu^\mp}\else{$\mu^\mp$}\fi}

 \def\nutau{\relax\ifmmode{\nu_\tau}\else{$\nu_\tau$}\fi}
 \def\nutaubar{\relax\ifmmode{\overline{\nu}\tiny_\tau}
	\else{$\overline{\nu}\tiny_\tau$}\fi}

 \def\tau{\relax\ifmmode{\mathchar"11C}\else{$\mathchar"11C$}\fi}
 \def\tauplus{\relax\ifmmode{\tau^+}\else{$\tau^+$}\fi}
 \def\tauminus{\relax\ifmmode{\tau^-}\else{$\tau^-$}\fi}
 \def\taupm{\relax\ifmmode{\tau^\pm}\else{$\tau^\pm$}\fi}
 \def\taump{\relax\ifmmode{\tau^\mp}\else{$\tau^\mp$}\fi}

 \def\nulep{\relax\ifmmode{\nu_\ell}\else{$\nu_\ell$}\fi}
 \def\nulepbar{\relax\ifmmode{\overline{\nu}\tiny_\ell}
	\else{$\overline{\nu}\tiny_\ell$}\fi}

 \def\lep{\relax\ifmmode{\ell}\else{$\ell$}\fi}
 \def\lepplus{\relax\ifmmode{\ell^+}\else{$\ell^+$}\fi}
 \def\lepminus{\relax\ifmmode{\ell^-}\else{$\ell^-$}\fi}
 \def\leppm{\relax\ifmmode{\ell^\pm}\else{$\ell^\pm$}\fi}
 \def\lepmp{\relax\ifmmode{\ell^\mp}\else{$\ell^\mp$}\fi}

 \def\pi{\relax\ifmmode{\mathchar"119}\else{$\mathchar"119$}\fi}
 \def\piplus{\relax\ifmmode{\pi^+}\else{$\pi^+$}\fi}
 \def\pizero{\relax\ifmmode{\pi^0}\else{$\pi^0$}\fi}
 \def\piminus{\relax\ifmmode{\pi^-}\else{$\pi^-$}\fi}
 \def\pipm{\relax\ifmmode{\pi^\pm}\else{$\pi^\pm$}\fi}
 \def\pimp{\relax\ifmmode{\pi^\mp}\else{$\pi^\mp$}\fi}
 \def\pipmz{\relax\ifmmode{\pi^{\pm,0}}\else{$\pi^{\pm,0}$}\fi}

 \def\eta{\relax\ifmmode{\mathchar"111}\else{$\mathchar"111$}\fi}
 \def\etazero{\relax\ifmmode{\eta^0}\else{$\eta^0$}\fi}

 \def\K{\relax\ifmmode{{\particletwo K}}\else{{\particletwo K}}\fi}
 \def\Kbar{\relax\ifmmode{\overline{{\particletwo K}}}
	\else{$\overline{{\particletwo K}}$}\fi}

 \def\Kzero{\relax\ifmmode{{\particletwo K}^0}\else{{\particletwo K}$^0$}\fi}
 \def\Kshort{\relax\ifmmode{{\particletwo K}^0_{\rm S}}
            \else{{\particletwo K}$^0_{\rm S}$}\fi}
 \def\Klong{\relax\ifmmode{{\particletwo K}^0_{\rm L}}
            \else{{\particletwo K}$^0_{\rm L}$}\fi}
 \def\Kzerobar{\relax\ifmmode{\overline{{\particletwo K}}\tiny^0}
	\else{$\overline{{\particletwo K}}\tiny^0$}\fi}

 \def\Kplus{\relax\ifmmode{{\particletwo K}^+}\else{{\particletwo K}$^+$}\fi}
 \def\Kminus{\relax\ifmmode{{\particletwo K}^-}\else{{\particletwo K}$^-$}\fi}
 \def\Kpm{\relax\ifmmode{{\particletwo K}^\pm}\else{{\particletwo K}$^\pm$}\fi}
 \def\Kmp{\relax\ifmmode{{\particletwo K}^\mp}\else{{\particletwo K}$^\mp$}\fi}

 \def\D{\relax\ifmmode{{\particletwo D}}\else{{\particletwo D}}\fi}
 \def\Dbar{\relax\ifmmode{\overline{{\particletwo D}}}
	\else{$\overline{{\particletwo D}}$}\fi}

 \def\Dzero{\relax\ifmmode{{\particletwo D}^0}\else{{\particletwo D}$^0$}\fi}
 \def\Dzerobar{\relax\ifmmode{\overline{{\particletwo D}}\tiny^0}
	\else{$\overline{{\particletwo D}}\tiny^0$}\fi}

 \def\Dplus{\relax\ifmmode{{\particletwo D}^+}\else{{\particletwo D}$^+$}\fi}
 \def\Dminus{\relax\ifmmode{{\particletwo D}^-}\else{{\particletwo D}$^-$}\fi}
 \def\Dpm{\relax\ifmmode{{\particletwo D}^\pm}\else{{\particletwo D}$^\pm$}\fi}
 \def\Dmp{\relax\ifmmode{{\particletwo D}^\mp}\else{{\particletwo D}$^\mp$}\fi}

 \def\F{\relax\ifmmode{{\particletwo F}}\else{{\particletwo F}}\fi}
 \def\Fplus{\relax\ifmmode{{\particletwo F}^+}\else{{\particletwo F}$^+$}\fi}
 \def\Fminus{\relax\ifmmode{{\particletwo F}^-}\else{{\particletwo F}$^-$}\fi}
 \def\Fpm{\relax\ifmmode{{\particletwo F}^\pm}\else{{\particletwo F}$^\pm$}\fi}
 \def\Fmp{\relax\ifmmode{{\particletwo F}^\mp}\else{{\particletwo F}$^\mp$}\fi}

 \def\DUS{\relax\ifmmode{{\particletwo D}_s}\else{{\particletwo D}$_s$}\fi}
 \def\DUSO{\relax\ifmmode{{\particletwo D}_{s1}}
      \else{{\particletwo D}$_{s1}$}\fi}
 \def\DUSP{\relax\ifmmode{{\particletwo D}_s^+}
      \else{{\particletwo D}$_s^+$}\fi}
 \def\DUSM{\relax\ifmmode{{\particletwo D}_s^-}
      \else{{\particletwo D}$_s^-$}\fi}
 \def\DUSPM{\relax\ifmmode{{\particletwo D}_s^\pm}
           \else{{\particletwo D}$_s^\pm$}\fi}
 \def\DUSMP{\relax\ifmmode{{\particletwo D}_s^\mp}
           \else{{\particletwo D}$_s^\mp$}\fi}

 \def\B{\relax\ifmmode{{\particletwo B}}\else{{\particletwo B}}\fi}
 \def\Bbar{\relax\ifmmode{\overline{{\particletwo B}}}
	\else{$\overline{{\particletwo B}}$}\fi}

 \def\Bzero{\relax\ifmmode{{\particletwo B}^0}\else{{\particletwo B}$^0$}\fi}
 \def\Bzerobar{\relax\ifmmode{\overline{{\particletwo B}}\tiny^0}
	\else{$\overline{{\particletwo B}}\tiny^0$}\fi}

 \def\Bplus{\relax\ifmmode{{\particletwo B}^+}\else{{\particletwo B}$^+$}\fi}
 \def\Bminus{\relax\ifmmode{{\particletwo B}^-}\else{{\particletwo B}$^-$}\fi}
 \def\Bpm{\relax\ifmmode{{\particletwo B}^\pm}\else{{\particletwo B}$^\pm$}\fi}
 \def\Bmp{\relax\ifmmode{{\particletwo B}^\mp}\else{{\particletwo B}$^\mp$}\fi}

 \def\pro{\relax\ifmmode{{\particletwo p}}\else{{\particletwo p}}\fi}
 \def\probar{\relax\ifmmode{\overline{{\particletwo p}}}
	\else{$\overline{{\particletwo p}}$}\fi}

 \def\neu{\relax\ifmmode{{\particletwo n}}\else{{\particletwo n}}\fi}
 \def\neubar{\relax\ifmmode{\overline{{\particletwo n}}}
	\else{$\overline{{\particletwo n}}$}\fi}

 \def\Lambda{\relax\ifmmode{\mathchar"7003}\else{$\mathchar"7003$}\fi}
 \def\Lambdabar{\relax\ifmmode{\overline{\Lambda}}
	\else{$\overline{\Lambda}$}\fi}
 \def\Lambdazero{\relax\ifmmode{\Lambda^0}\else{$\Lambda^0$}\fi}
 \def\Lambdazerobar{\relax\ifmmode{\overline{\Lambda}\tiny^0}
	\else{$\overline{\Lambda}\tiny^0$}\fi}

 \def\Sigma{\relax\ifmmode{\mathchar"7006}\else{$\mathchar"7006$}\fi}
 \def\Sigmabar{\relax\ifmmode{\overline{\Sigma}}
	\else{$\overline{\Sigma}$}\fi}
 \def\Sigmazero{\relax\ifmmode{\Sigma^0}\else{$\Sigma^0$}\fi}
 \def\Sigmazerobar{\relax\ifmmode{\overline{\Sigma}\tiny^0}
	\else{$\overline{\Sigma}\tiny^0$}\fi}
 \def\Sigmaplus{\relax\ifmmode{\Sigma^+}\else{$\Sigma^+$}\fi}
 \def\Sigmaminus{\relax\ifmmode{\Sigma^-}\else{$\Sigma^-$}\fi}
 \def\Sigmapm{\relax\ifmmode{\Sigma^\pm}\else{$\Sigma^\pm$}\fi}
 \def\Sigmamp{\relax\ifmmode{\Sigma^\mp}\else{$\Sigma^\mp$}\fi}
 \def\Sigmaplusbar{\relax\ifmmode{\overline{\Sigma}\tiny^+}
	\else{$\overline{\Sigma}\tiny^+$}\fi}
 \def\Sigmaminusbar{\relax\ifmmode{\overline{\Sigma}\tiny^-}
	\else{$\overline{\Sigma}\tiny^-$}\fi}
 \def\Sigmapmbar{\relax\ifmmode{\overline{\Sigma}\tiny^\pm}
	\else{$\overline{\Sigma}\tiny^\pm$}\fi}
 \def\Sigmampbar{\relax\ifmmode{\overline{\Sigma}\tiny^\mp}
	\else{$\overline{\Sigma}\tiny^\mp$}\fi}

 \def\Xi{\relax\ifmmode{\mathchar"7004}\else{$\mathchar"7004$}\fi}
 \def\Xibar{\relax\ifmmode{\overline{\Xi}}
	\else{$\overline{\Xi}$}\fi}
 \def\Xizero{\relax\ifmmode{\Xi^0}\else{$\Xi^0$}\fi}
 \def\Xizerobar{\relax\ifmmode{\overline{\Xi}\tiny^0}
	\else{$\overline{\Xi}\tiny^0$}\fi}
 \def\Ximinus{\relax\ifmmode{\Xi^-}\else{$\Xi^-$}\fi}
 \def\Xiplusbar{\relax\ifmmode{\overline{\Xi}\tiny^+}
	\else{$\overline{\Xi}\tiny^+$}\fi}

 \def\Omegaminus{\relax\ifmmode{\Omega^-}\else{$\Omega^-$}\fi}
 \def\Omegaplusbar{\relax\ifmmode{\overline{\Omega}\tiny^+}
	\else{$\overline{\Omega}\tiny^+$}\fi}

 \def\Lambdac{\relax\ifmmode{\Lambda_c}\else{$\Lambda_c$}\fi}
 \def\Lambdacbar{\relax\ifmmode{\overline{\Lambda_c}}
	\else{$\overline{\Lambda}\tiny_c$}\fi}
 \def\Lambdacplus{\relax\ifmmode{\Lambda^+_c}\else{$\Lambda^+_c$}\fi}
 \def\Lambdacminusbar{\relax\ifmmode{\overline{\Lambda}\tiny^-_c}
	\else{$\overline{\Lambda}\tiny^-_c$}\fi}

 \def\rho{\relax\ifmmode{\mathchar"11A}\else{$\mathchar"11A$}\fi}
 \def\rhoplus{\relax\ifmmode{\rho^+}\else{$\rho^+$}\fi}
 \def\rhozero{\relax\ifmmode{\rho^0}\else{$\rho^0$}\fi}
 \def\rhominus{\relax\ifmmode{\rho^-}\else{$\rho^-$}\fi}
 \def\rhopm{\relax\ifmmode{\rho^\pm}\else{$\rho^\pm$}\fi}
 \def\rhomp{\relax\ifmmode{\rho^\mp}\else{$\rho^\mp$}\fi}
 \def\rhopmz{\relax\ifmmode{\rho^{\pm,0}}\else{$\rho^{\pm,0}$}\fi}

 \def\omega{\relax\ifmmode{\mathchar"121}\else{$\mathchar"121$}\fi}
 \def\omegazero{\relax\ifmmode{\omega^0}\else{$\omega^0$}\fi}

 \def\etaprime{\relax\ifmmode{\eta'}\else{$\eta'$}\fi}
 \def\etaprimezero{\relax\ifmmode{\eta'\tiny^0}\else{$\eta'\tiny^0$}\fi}

 \def\phi{\relax\ifmmode{\mathchar"11E}\else{$\mathchar"11E$}\fi}

 \def\psi{\relax\ifmmode{\mathchar"120}\else{$\mathchar"120$}\fi}
 \def\psiprime{\relax\ifmmode{\psi'}\else{$\psi'$}\fi}
 \def\psidoubleprime{\relax\ifmmode{\psi''}\else{$\psi''$}\fi}
 \def\psitripleprime{\relax\ifmmode{\psi'''}\else{$\psi'''$}\fi}
 \def\psifourprime{\relax\ifmmode{\psi''''}\else{$\psi''''$}\fi}

 \def\Uones{\relax\ifmmode{\Upsilon{\particletwo (1S)}}
	\else{$\Upsilon{\particletwo (1S)}$}\fi}
 \def\Utwos{\relax\ifmmode{\Upsilon{\particletwo (2S)}}
	\else{$\Upsilon{\particletwo (2S)}$}\fi}
 \def\Uthrees{\relax\ifmmode{\Upsilon{\particletwo (3S)}}
	\else{$\Upsilon{\particletwo (3S)}$}\fi}
 \def\Ufours{\relax\ifmmode{\Upsilon{\particletwo (4S)}}
	\else{$\Upsilon{\particletwo (4S)}$}\fi}
 \def\Ufives{\relax\ifmmode{\Upsilon{\particletwo (5S)}}
	\else{$\Upsilon{\particletwo (5S)}$}\fi}
 \def\Usixs{\relax\ifmmode{\Upsilon{\particletwo (6S)}}
	\else{$\Upsilon{\particletwo (6S)}$}\fi}

 \def\Kstar{\relax\ifmmode{{\particletwo K}^\star}
\else{{\particletwo K}$^\star$}\fi}
 \def\Kstarbar{\relax\ifmmode{\overline{{\particletwo K}}\tiny^\star}
	\else{$\overline{{\particletwo K}}\tiny^\star$}\fi}

 \def\Kstarzero{\relax\ifmmode{{\particletwo K}^{\star0}}
\else{{\particletwo K}$^{\star0}$}\fi}
 \def\Kstarzerobar{\relax\ifmmode{\overline{{\particletwo K}}\tiny^{\star0}}
	\else{$\overline{{\particletwo K}}\tiny^{\star0}$}\fi}

 \def\Kstarplus{\relax\ifmmode{{\particletwo K}^{\star+}}
\else{{\particletwo K}$^{\star+}$}\fi}
 \def\Kstarminus{\relax\ifmmode{{\particletwo K}^{\star-}}
\else{{\particletwo K}$^{\star-}$}\fi}
 \def\Kstarpm{\relax\ifmmode{{\particletwo K}^{\star\pm}}
\else{{\particletwo K}$^{\star\pm}$}\fi}
 \def\Kstarmp{\relax\ifmmode{{\particletwo K}^{\star\mp}}
\else{{\particletwo K}$^{\star\mp}$}\fi}

 \def\Dstar{\relax\ifmmode{{\particletwo D}^\star}
\else{{\particletwo D}$^\star$}\fi}
 \def\Dstarbar{\relax\ifmmode{\overline{{\particletwo D}}\tiny^\star}
	\else{$\overline{{\particletwo D}}\tiny^\star$}\fi}

 \def\Dstarzero{\relax\ifmmode{{\particletwo D}^{\star0}}
\else{{\particletwo D}$^{\star0}$}\fi}
 \def\Dstarzerobar{\relax\ifmmode{\overline{{\particletwo D}}\tiny^{\star0}}
	\else{$\overline{{\particletwo D}}\tiny^{\star0}$}\fi}

 \def\Dstarplus{\relax\ifmmode{{\particletwo D}^{\star+}}
\else{{\particletwo D}$^{\star+}$}\fi}
 \def\Dstarminus{\relax\ifmmode{{\particletwo D}^{\star-}}
\else{{\particletwo D}$^{\star-}$}\fi}
 \def\Dstarpm{\relax\ifmmode{{\particletwo D}^{\star\pm}}
\else{{\particletwo D}$^{\star\pm}$}\fi}
 \def\Dstarmp{\relax\ifmmode{{\particletwo D}^{\star\mp}}
\else{{\particletwo D}$^{\star\mp}$}\fi}

 \def\Ddoublestar{\relax\ifmmode{{\particletwo D}^{\star\star}}
\else{{\particletwo D}$^{\star\star}$}\fi}
 \def\Ddoublestarbar{\relax\ifmmode{\overline{{\particletwo D}}
\tiny^{\star\star}}
	\else{$\overline{{\particletwo D}}\tiny^{\star\star}$}\fi}

 \def\Ddoublestarzero{\relax\ifmmode{{\particletwo D}^{\star\star0}}
	\else{{\particletwo D}$^{\star\star0}$}\fi}
 \def\Ddoublestarzerobar{\relax\ifmmode
{\overline{{\particletwo D}}\tiny^{\star\star0}}
	\else{$\overline{{\particletwo D}}\tiny^{\star\star0}$}\fi}

 \def\Ddoublestarplus{\relax\ifmmode{{\particletwo D}^{\star\star+}}
	\else{{\particletwo D}$^{\star\star+}$}\fi}
 \def\Ddoublestarminus{\relax\ifmmode{{\particletwo D}^{\star\star-}}
	\else{{\particletwo D}$^{\star\star-}$}\fi}
 \def\Ddoublestarpm{\relax\ifmmode{{\particletwo D}^{\star\star\pm}}
	\else{{\particletwo D}$^{\star\star\pm}$}\fi}
 \def\Ddoublestarmp{\relax\ifmmode{{\particletwo D}^{\star\star\mp}}
	\else{{\particletwo D}$^{\star\star\mp}$}\fi}

 \def\Bstar{\relax\ifmmode{{\particletwo B}^\star}
\else{{\particletwo B}$^\star$}\fi}
 \def\Bstarbar{\relax\ifmmode{\overline{{\particletwo B}}\tiny^\star}
	\else{$\overline{{\particletwo B}}\tiny^\star$}\fi}

 \def\Bstarzero{\relax\ifmmode{{\particletwo B}^{\star0}}
\else{{\particletwo B}$^{\star0}$}\fi}
 \def\Bstarzerobar{\relax\ifmmode{\overline{{\particletwo B}}\tiny^{\star0}}
	\else{$\overline{{\particletwo B}}\tiny^{\star0}$}\fi}

 \def\Bstarplus{\relax\ifmmode{{\particletwo B}^{\star+}}
\else{{\particletwo B}$^{\star+}$}\fi}
 \def\Bstarminus{\relax\ifmmode{{\particletwo B}^{\star-}}
\else{{\particletwo B}$^{\star-}$}\fi}
 \def\Bstarpm{\relax\ifmmode{{\particletwo B}^{\star\pm}}
\else{{\particletwo B}$^{\star\pm}$}\fi}
 \def\Bstarmp{\relax\ifmmode{{\particletwo B}^{\star\mp}}
\else{{\particletwo B}$^{\star\mp}$}\fi}

 \def\Vud{\relax\ifmmode{{\rm V}_{ud}}\else{{\rm V}$_{ud}$}\fi}
 \def\Vcd{\relax\ifmmode{{\rm V}_{cd}}\else{{\rm V}$_{cd}$}\fi}
 \def\Vtd{\relax\ifmmode{{\rm V}_{td}}\else{{\rm V}$_{td}$}\fi}
 \def\Vus{\relax\ifmmode{{\rm V}_{us}}\else{{\rm V}$_{us}$}\fi}
 \def\Vcs{\relax\ifmmode{{\rm V}_{cs}}\else{{\rm V}$_{cs}$}\fi}
 \def\Vts{\relax\ifmmode{{\rm V}_{ts}}\else{{\rm V}$_{ts}$}\fi}
 \def\Vub{\relax\ifmmode{{\rm V}_{ub}}\else{{\rm V}$_{ub}$}\fi}
 \def\Vcb{\relax\ifmmode{{\rm V}_{cb}}\else{{\rm V}$_{cb}$}\fi}
 \def\Vtb{\relax\ifmmode{{\rm V}_{tb}}\else{{\rm V}$_{tb}$}\fi}
 \def\u{\relax\ifmmode{{\rm u}}\else{{\rm u}}\fi}

}
\conventionone
\begin{document}
\title{Heavy Quark Hadronic Weak Decays from CLEO-II}
\author{Harry N. Nelson\footnotemark[1]\ \\
       {\it Department of Physics} \\
       {\it University of California } \\
       {\it Santa Barbara, California 93106-9530, USA}}

\abstract{\rightskip=1.5pc
          \leftskip=1.5pc
We present preliminary results from the CLEO-II collaboration
on a variety of hadronic final states of mesons
containing heavy quarks.  In particular, the
pattern of 2-body \B\ decays is now decisively
different that that of \D\ and \K\ decays;
perhaps a consequence will be that
$\tau_{\BM}<\tau_{\BZB}$.
We have
observed `wrong-sign' $\DZ\sto\KP\PIM$ decays,
which are probably due to a doubly Cabibbo-suppressed
transition.}

\maketitle

\def\thefootnote{\fnsymbol{footnote}}
\footnotetext[1]{Invited Talk Presented
at the International Europhysics Conference on {\bf High Energy
Physics}, Marseille, July 22 - July 28, 1993; supported by the
U.S. Department of Energy Award No. DE--FG03--92--ER40618-L.}
\vspace{2.4mm}
{\bf \noindent \underline{Introduction}}
\vspace{0.9mm}
\par
\nobreak
We present preliminary results on the decays
of mesons containing $b$ and $c$ quarks to
a variety of hadronic final states.   The data
sample for these analyses is typically
$1-1.8{\rm fb}^{-1}$ accumulated by the CLEO-II
detector, 2/3 of which is on the \USSSS,
and 1/3 of which is in the continuum just below that
resonance.

 Of primary importance are the high statistics
measurements of the decays of \B-mesons to two
body final states.  The evidence is now 
that the description of decays to two body final states for
\D\ and \K\ mesons {\it does not pertain for\/} \B's.

 We present results on inclusive
measurements of \D\ and \PS\ mesons in \B\ decays.
Also, results on Cabibbo-suppressed decays of \D\ mesons,
the observation of the wrong-sign decay $\DZ\sto\KP\PIM$,
and precision measurement of absolute branching ratios for
the \DZ\ and \DP\ are presented.
 
\vspace{2.3mm}
{\bf \noindent \underline{\boldmath Two Body Decays of \B's}}
\vspace{0.9mm}
\par
\begin{figure}[ht]
\hbox{\hfill\hskip0.1in\epsffile{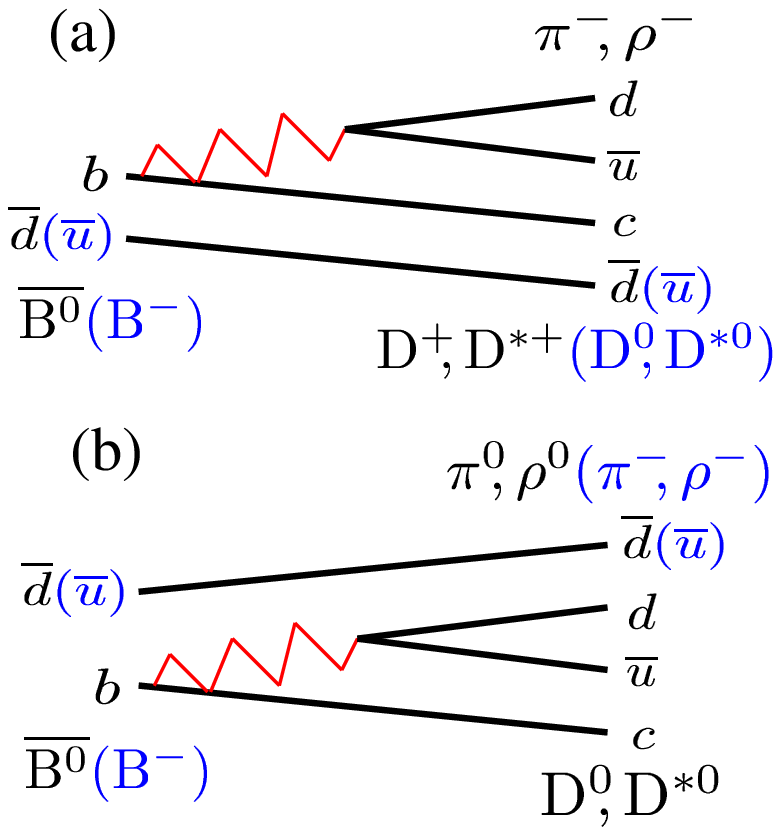}\hfill} 
\caption{Spectator Diagrams for Two-Body \B\ Decay:
(a) External, which under the assumption of factorization
can be related to an exclusive semileptonic amplitude,
and (b) Internal, which can suffer color suppression.
For the \BM, the two amplitudes add {\it constructively},
according the results presented here.}
\label{fig:spectators}
\end{figure}

\nobreak
Consider the generic Cabibbo-favored
decay of a flavored pseudoscalar
meson with one light quark,
\eM, to two pseudoscalars.
The neutral pseudoscalar, \eMZ, can
decay either to a charged final state $-+$
or a neutral final state $(00)$; the charged
pseudoscalar, \eMP, decays to a mixed charge
final state, $0+$; consider 
$\Gamma(\eMZ\sto-+)/\Gamma(\eMP\to0+)$.  In the
kaon system, due to the \diha\ rule:
\begin{equation}
\Gamma(\KZB\sto\PIP\PIM)/\Gamma(\KM\sto\PIM\PIZ)\approx225
\end{equation}
In the \D\ system, where the relative enhancement of the
amplitude that produces the smallest change in isospin
is not as prominent,
\begin{equation}
\Gamma(\DZ\sto\KM\PIP)/\Gamma(\DP\sto\KZB\PIP)\approx3.6
\end{equation}
However, the results we present here indicate, assuming
$\tau_{\BM}=\tau_{\BZB}$, and that the
\USSSS\ decays equally to \BP\BM\
and \BZ\BZB\ $(f_{+-}=f_{00})$,
\begin{equation}
\Gamma(\BZB\sto\DP\PIM)/\Gamma(\BM\sto\DZ\PIM)\approx0.6\pm0.1
\end{equation}
decisively {\it less\/} than unity.  Because the ratio
$\Gamma(-+)/\Gamma(0+)$ is near to unity,
the spectator quark 
processes shown in Fig.~\ref{fig:spectators}, rather
than processes with more complicated light quark
interactions in the final state,
presumably dominate the two-body
decay amplitudes for the \B\ system.  Because
$\Gamma(-+)/\Gamma(0+)<1$, the two
amplitudes for \BM\ decay that lead to identical
final states add {\it constructively}.

\epsfxsize=3.0in
\vspace{2.3mm}

\begin{table}[ht]
\begintable
            Mode         | ${\cal B}$ Assumed     | Source \cr
$\DZ\sto\KM\PIP$         | $(3.91\pm0.10)\%$ | CLEO-II \nr
$\DZ\sto\KM\PIP\PIZ$      | $(12.1\pm1.1)\%$  | PDG    \nr
$\DZ\sto\KM\PIP\PIP\PIM$ | $(8.0\pm0.5)\%$   | PDG     \nr 
$\DP\sto\KM\PIP\PIP$     | $(10.0\pm1.4)\%$  | CLEO-II \cr
$\DSP\sto\DZ\PIP$        | $(67.9\pm2.3)\%$  | CLEO-II \nr
$\DSZ\sto\DZ\PIZ$        | $(62.5\pm4.2)\%$  | CLEO-II
\endtable
\caption{ Charm Decay Modes and Branching Ratios
used in the \B\ reconstructions.}
\label{tab:dmodes}
\end{table}
\vspace{2.3mm}

The reconstructed \D\ decay modes are shown in Table~\ref{tab:dmodes}.
Two variables, the beam constrained mass, 
$M^2_{\B}=E^2_{\rm beam}-(\sum_i\vec{p_i})^2$, and the
energy difference with the beam, $\Delta E =E_{\rm beam}-(\sum_i E_i)$
where the sums run over the particles assigned to the $\D(\overline{u}d)$
system, are important in the \B\ reconstruction.  Typically
$\sigma_{M_{\B}}\approx2.6\,{\rm MeV}$, and is insensitive to the
final state mode, while $\sigma_{\Delta E}\approx15-40\,{\rm MeV}$
and is sensitive to the final state mode. 

\epsfxsize=3.0in
\begin{figure}[ht]
\hbox{\hfill\hskip0.1in\epsffile{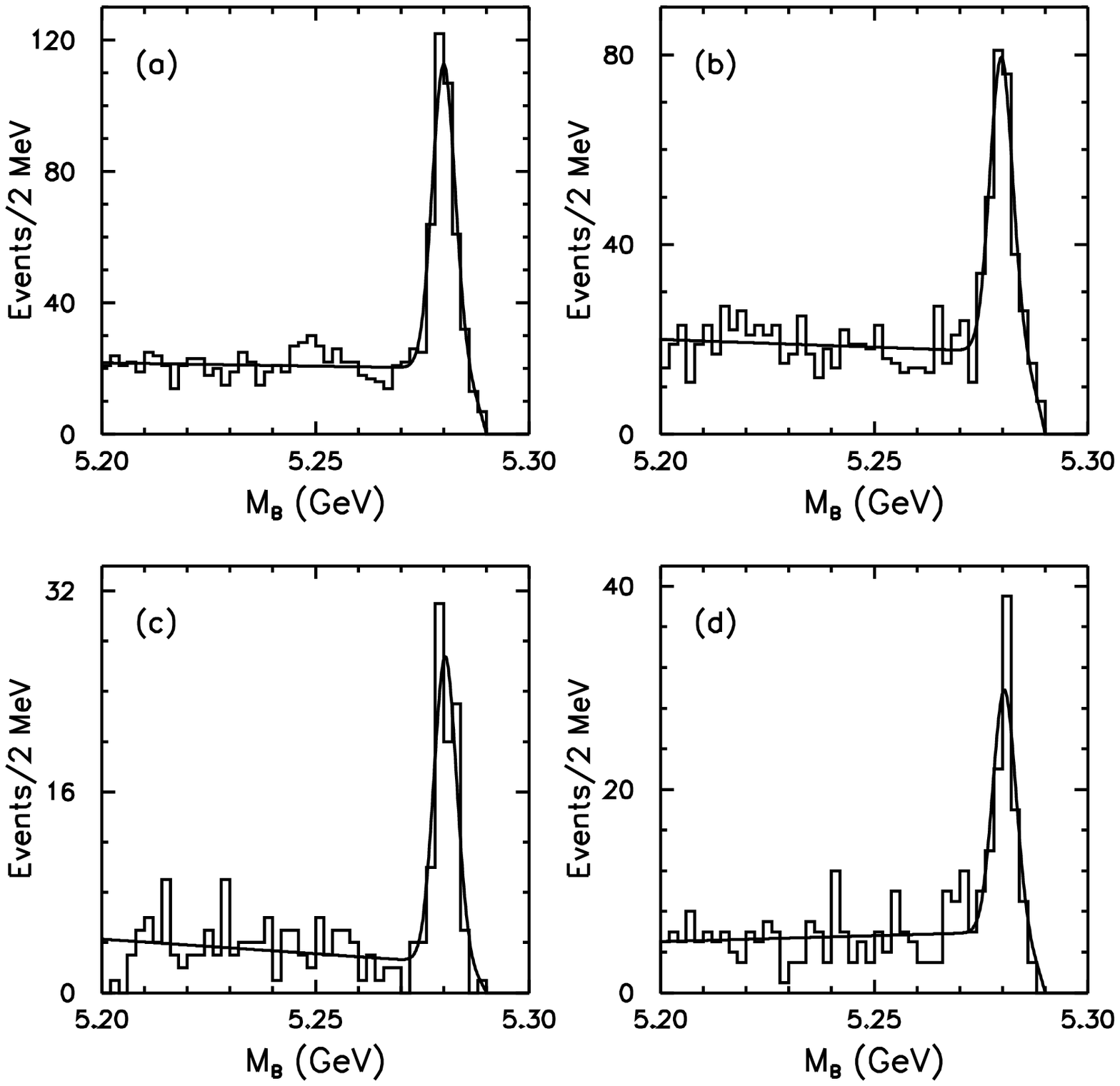}\hfill} 
\caption{Beam constrained mass ($M_{\B}$) distributions for
(a) $\BM\sto\DZ\PIM$ decays; 
(b) $\BM\sto\DZ\RHM$ decays;
(c) $\BZB\sto\DP\PIM$ decays;
and (d) $\BZB\sto\DP\RHM$ decays.}
\label{fig:typsig}
\end{figure}
\vspace{2.3mm}

Typical signals in the $\B\sto\D(\PI{\rm or}\RH)$ modes are
shown in Fig.~\ref{fig:typsig}.  The background function
is determined in several ways, including studies of the
sidebands and Monte Carlo simulations: the background function
shape is linear far from the \B\ mass, and parabolic just
under the \B\ mass.
Branching ratio results for $+-$ and $0-$ modes are given
in Tables~\ref{tab:resone} and~\ref{tab:restwo}, in which
the assumption that the \USSSS\ decays equally to \BP\BM\
and \BZ\BZB\ $(f_{+-}=f_{00})$ is made.

\begin{table}[ht]
\begintable
 \BZB$\sto$ & \# & ${\cal B}(\%)$ \cr 
 \DP\PIM & 76\PM10 & $0.22^{\PM 0.03}_{\PM 0.02 \PM 0.03}$  \cr
\DSP\PIM & 73\PM10 & $0.27^{\PM 0.04}_{\PM 0.04 \PM 0.013}$ \cr
 \DP\RHM & 86\PM11 & $0.62^{\PM 0.08}_{\PM 0.08 \PM 0.09}$  \cr
\DSP\RHM & 52\PM8  & $0.74^{\PM 0.11}_{\PM 0.13 \PM 0.03}$
\endtable
\caption{Two body decay modes of the \BZB\@. The top error
is statistical, on the inner bottom is intrinsic systematic
error, and on the outer bottom is the extrinsic systematic
error from, for example, errors in \D\ branching ratios.}
\label{tab:resone}
\end{table}
\vspace{2.3mm}

\begin{table}[ht]
\begintable
 \BM$\sto$  & \# & ${\cal B}(\%)$ \cr     
\DZ\PIM  & 302\PM22 & $0.47^{\PM 0.03}_{\PM 0.05 \PM 0.02}$  \cr
\DSZ\PIM & 93\PM12  & $0.50^{\PM 0.06}_{\PM 0.07 \PM 0.04}$  \cr
\DZ\RHM  & 248\PM22 & $1.07^{\PM 0.10}_{\PM 0.16 \PM 0.04}$  \cr
\DSZ\RHM & 92\PM12  & $1.41^{\PM 0.19}_{\PM 0.13 \PM 0.11}$
\endtable
\caption{Two body decay modes of the \BM\@.  The error notation
is the same as the previous table.}
\label{tab:restwo}
\end{table}
\vspace{8mm}

One can see in all cases that ${\cal B}(\BM\sto0-)>{\cal B}(\BZB\sto+-)$.
One simple physical explanation is that the diagrams of Fig.~1(a) and
Fig.~1(b) add constructively for the \BM\@.

\begin{figure}[ht]
\hbox{\hfill\hskip0.1in\epsffile{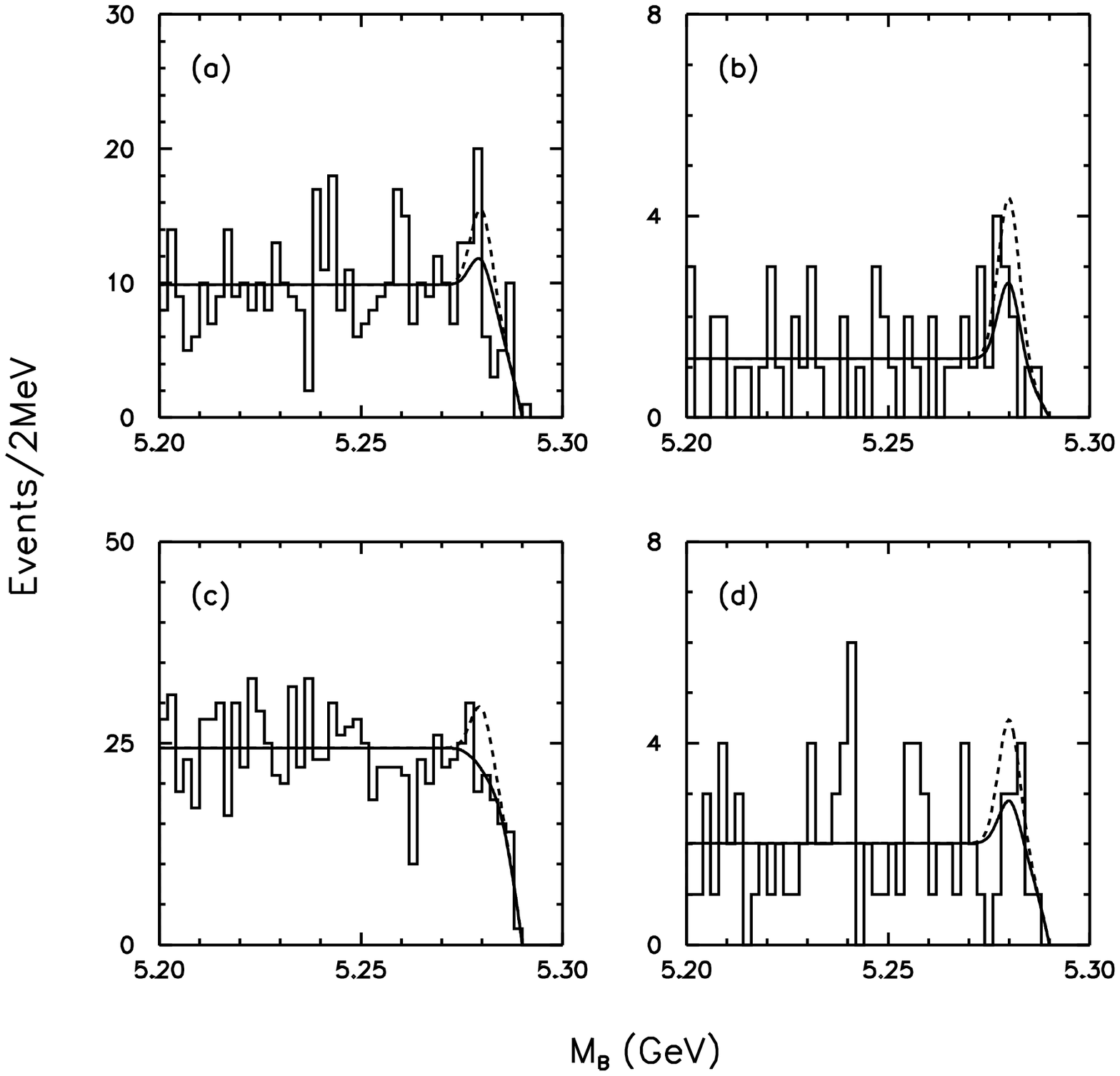}\hfill} 
\caption{Beam constrained mass ($M_{\B}$) distributions for
(a) $\BZB\sto\DZ\PIZ$ decays; 
(b) $\BZB\sto\DSZ\PIZ$ decays;
(c) $\BZB\sto\DZ\RHZ$ decays;
and (d) $\BZB\sto\DSZ\RHZ$ decays.  The solid curves show just the
background shape, the dotted curves the 90\% CL upper limit.}
\label{fig:aytoo}
\end{figure}

No evidence exists in our data sample for the mode $\BZB\sto00$.
The plots for $M_{\B}$ for the various modes are shown in
Fig.~\ref{fig:aytoo}.  Based upon the absence of signal in these
plots, one arrives at the limits in Table~\ref{tab:aytoo}.  Note
that the ratio $\Gamma(00)/\Gamma(+-)$ is at least less than
$1/4$ for the \BZB, in marked contrast to the situation for the
\DZ and the \KZB, where this ratio is typically $1/2$.

\epsfxsize=3.0in
\vspace{2.3mm}

\begin{table}[ht]
\begintable
 \BZB$\rightarrow$ & ${\cal B}\%$ (90\% C.L.) & %
$\Gamma(00)\over{\Gamma(+-)}$ \% & BSW (\%) \cr
 \DZ\PIZ & $<0.03$ & $<14$ & $76\vert{a_2\over a_1}\vert^2$ \cr
 \DSZ\PIZ & $<0.06$ & $<22$ & 
 $84\vert{a_2\over a_1}\vert^2$ \cr
 \DZ\RHZ & $<0.08$ & $<13$ & 
 $22\vert{a_2\over a_1}\vert^2$ \cr
 \DSZ\RHZ & $<0.17$ & $<23$ & 
$32\vert{a_2\over a_1}\vert^2$ \endtable
\caption{Limits on decays of the type $\BZ\sto00$, which
can proceed via the internal spectator amplitude of Fig.~1(b).
In addition to $f_{+-}=f_{00}$, $\tau_{\BM}=\tau_{\BZB}$ is assumed for
all extractions of the BSW parameters $a_1$ and $a_2$.}
\label{tab:aytoo}
\end{table}
\vspace{2.3mm}

The two body decay data can be described by the phenomenology
of Bauer, Stech, and Wirbel (BSW)[1], where the external spectator in 
Fig.~\ref{fig:spectators}(a) is associated with the coefficient
$a_1$, and the internal spectator in Fig.~\ref{fig:spectators}(b)
is associated with the coefficient $a_2$.  From the decays
$\BZB\sto+-$, which are purely external spectator, one can
infer $|a_1|=0.98\pm0.03\pm0.04\pm0.09$, as detailed in
Table~\ref{tab:ayone}.  One can also test the correctness
of the association of $\BZB\sto+-$ with the external spectator
by relating its branching ratio to $\B\sto\D\ell\nu$, an
association known also as factorization.  The physical content
is simply the replacement of the hadronization of the 
$W^-\sto\overline{u}d\sto\pi^-$ with $W^-\sto\ell^-\overline{\nu}$.
Factorization predicts ${\cal B}(\BZB\sto\DSP\PIM)=
6\pi^2{c_1^2} f_{\pi}^2\vert V_{ud}\vert^2\!\times\! 
{{d {\cal B}}\over{dQ^2}}(B\!\to\!D^* \L~\nu)\vert_{Q^2=m_{\pi}^2} 
=(0.26\PM0.04\%)$, and ${\cal B}(\BZB\sto\DSP\RHM)=
6\pi^2{c_1^2} f_{\rho}^2\vert V_{ud}\vert^2\!\times\! 
{{d {\cal B}}\over{dQ^2}}(B\!\to D^*\!\L~\nu)\vert_{Q^2=m_{\rho}^2}$=
$(0.75\PM0.10\%)$, in good agreement with the measurements.

\begin{table}[ht]
\begintable
 \BZB$\to$ & ${\cal B}$(\%) & BSW \cr 
\DP\PIM  & 0.22\PM0.05  & 0.264$\vert a_1\vert^2$    \nr
\DSP\PIM & 0.27\PM0.05  & 0.254$\vert a_1\vert^2$    \nr
\DP\RHM  & 0.62\PM0.14  & 0.621$\vert a_1\vert^2$    \nr
\DSP\RHM & 0.74\PM0.17  & 0.702$\vert a_1\vert^2$
\endtable
\caption{Comparison of measured $\BZB\sto+-$ branching ratios
with the BSW parameterization}
\label{tab:ayone}
\end{table}

We see, from
Table~\ref{tab:aytoo}, that the absence of $\BZB\sto00$ modes
imply that $|a_2|<0.5$ or so.  There are two ways in which we
obtain increased sensitivity to $a_2$: first, for the \BM\ decays,
in the BSW phenomenology, the external and internal amplitudes
coherently interfere, so the rates for $\BM\sto0-$ are crudely
$\propto|a_1+a_2|^2$, yielding a linear sensitivity to $a_2$ in the
interference term; second, we can measure the modes produced
by the internal spectator diagram where the \W\ hadronizes as
a $\overline{c}s$ rather than $\overline{u}d$, such as
$\BZB\sto\PS\KSH$.  For the first method, define:
\begin{eqnarray}
R_1 & = {{\cal B}(\BZB\to\DP\PIM) \over {\cal B}(\BM\to\DZ\PIM)}
                & = {1\over{(1 + 1.23 a_2/a_1)^2}}    \\
R_2 & = {{\cal B}(\BZB\to\DSM\PIM) \over {\cal B}(\BM\to\DSZ\PIM)}
                & = {1\over{(1 + 1.292 a_2/a_1)^2}}   \\
R_3 & = {{\cal B}(\BZB\to\DP\RHM) \over {\cal B}(\BM\to \DZ\RHM) }
                & = {1\over{(1 + 0.662 a_2 /a_1)^2}}  \\
R_4 & = {{\cal B}(\BZB\to\DSM\RHM) \over {\cal B}(\BM\to\DSZ\RHM)}
                & \approx {1\over{(1 + 1.5 a_2/a_1)^2}}
\end{eqnarray}

With these definitions, we find the results given in 
Table~\ref{tab:aytwoint}, which indicate $a_2/a_1\approx0.24$.
Note the relative sign is {\it positive}, in contradiction to
the destructive interference obtained in the BSW analysis of the
analogous charm decays.

\begin{table}[ht]
\begintable
Ratio &${a_2\over a_1} =-0.24 $ & ${a_2\over a_1} =0.24 $ & {\bf CLEO-II} \cr
$R_1 $& 2.0  & 0.59 & $ 0.56 \pm 0.09 \pm 0.11$ \nr
$R_2 $& 2.1  & 0.58 & $ 0.64 \pm 0.06 \pm 0.05$ \nr
$R_3 $& 1.4  & 0.74 & $ 0.69 \pm 0.11 \pm 0.12$ \nr
$R_4 $& 1.3  & 0.54 & $ 0.63 \pm 0.07 \pm 0.05$ 
\endtable
\caption{Estimation of $a_2$ by interference in $\BM\sto0-$
decays.}
\label{tab:aytwoint}
\end{table}

\begin{figure}[ht]
\hbox{\hfill\hskip0.1in\epsffile{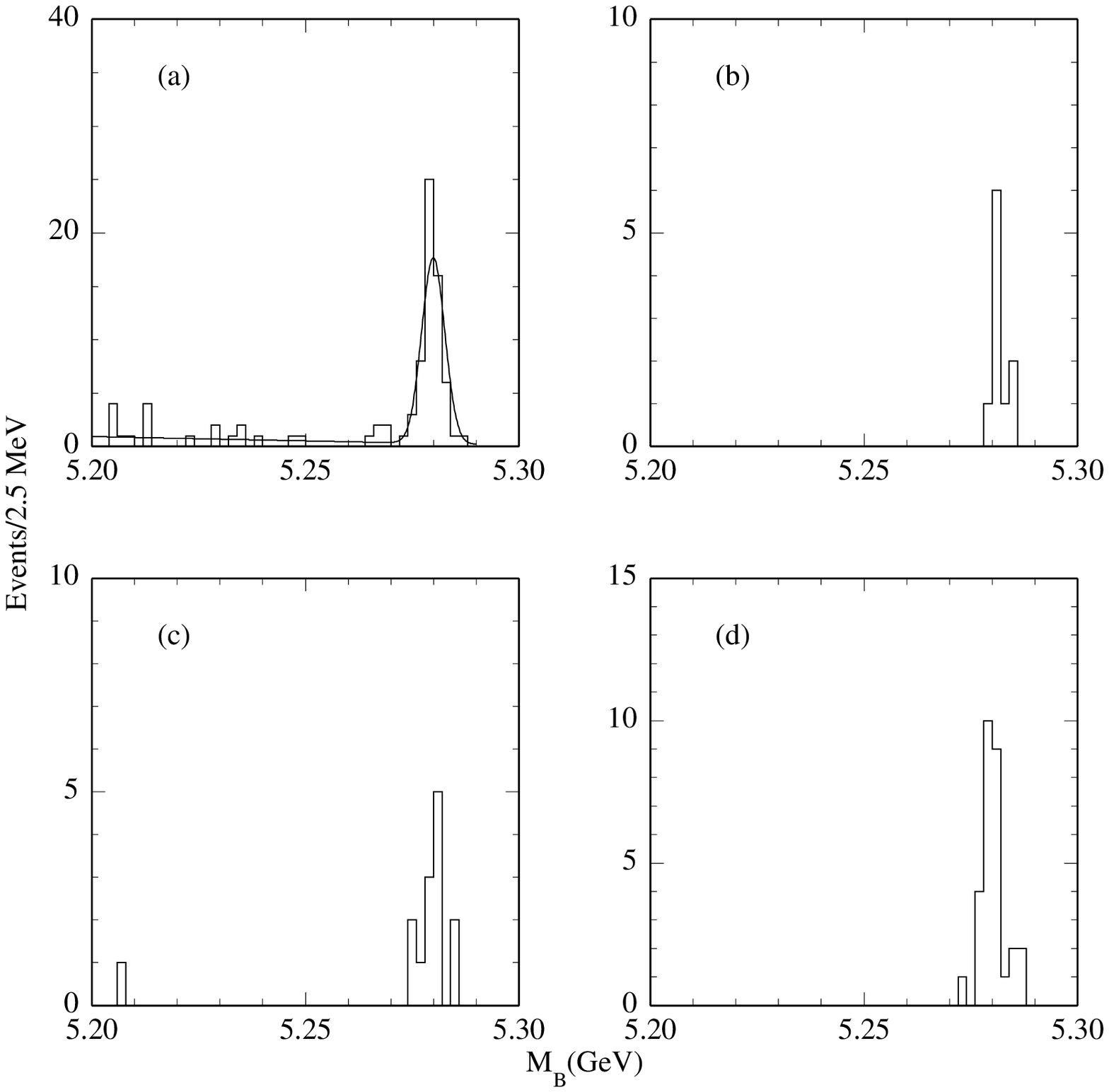}\hfill}
\caption{Beam constrained mass ($M_{\B}$) distributions for
(a) $\BM\sto\PS\KM$ decays; 
(b) $\BZB\sto\PS\KSH$ decays;
(c) $\BM\sto\PS\KSM$ decays;
and (d) $\BZB\sto\PS\KSZB$ decays.}
\label{fig:kjpsi}
\end{figure}

When the \W\ hadronizes as $\overline{c}s$, the internal
spectator can produce the decays $\B\sto\PS\K$.  The decays
of the \BZB\ of this type produce CP eigenstates, and are
expected to be useful in the measurement of CP violation in
the $\BZ\!-\!\BZB$ system, in particular to extract $\sin{2\beta}$.
The CLEO-II signals in these modes, where the $\PS\sto\ell^+\ell^-$,
are shown in Fig.~\ref{fig:kjpsi}.  The numbers for extraction
of $a_2$ are shown in Table~\ref{tab:kjpsi}, and yield
$|a_2|=0.25\pm0.013\pm0.006\pm0.02$, in agreement with the
determination from interference.

\epsfxsize=3.0in
\vspace{2.3mm}

\begin{table}[ht]
\begintable
 \BZB$\to$ & ${\cal B}(\%)$ | BSW (\%) | ${\cal B}(\%)$  &
\BM$\rightarrow$ \cr      
 \PS\KSH & $0.08^{\PM0.03}_{\PM0.01}$ | $1.82\vert a_2\vert^2$ |
                 $0.11^{\PM0.02}_{\PM0.01}$ & \PS\KM \nr
 \PS\KSZB & $0.19^{\PM0.04}_{\PM0.02}$ | $2.93\vert a_2\vert^2$ |
                 $0.21^{\PM0.06}_{\PM0.03}$ &  \PS\KSM
\endtable
\caption{ Measurement of $|a_2|$ by rate of $\B\sto\PS\K$ decays.}
\label{tab:kjpsi}
\end{table}

To conclude this discussion of the two body decays of the \B:
given that a number of branching ratios for the \BM\ are greater
than those for the \BZB, one can wonder whether the
oft-quoted prediction that $\tau_{\BM}>\tau_{\BZB}$ really has a
solid foundation.  One can see that differences
between exclusive $\BM\sto0-$ and $\BZB\sto+-$ partial rates where the
\W\ hadronizes as $\overline{u}d$, will wash out in the inclusive
decay rate: partial widths when the \W\ hadronizes as $\overline{c}s$
or couples to leptons are surely the same between \BM\ and \BZB\@.
What is hard to see is how the remaining decay rates, predominantly
high multiplicity decays where the \W\ hadronizes $\overline{u}d$,
could push the inclusive \BZB\ decay rate higher than the \BM\@.

{\bf \noindent\
\underline{\boldmath Inclusive measurements of \D's and $J/\psi$'s}}
\vspace{0.9mm}
\par
\nobreak
We have recently made new measurements of the inclusive branching
ratios of \B\ mesons to various openly charmed and hidden charmed
mesons.  The statistics involved in these measurements is
much better than earlier results: for example, about 1500 events
are used to measure $\B\sto\PS X$.
These measurements are summarized in Table~\ref{tab:inclus}.
Whether the excess of \DSZ\ relative to \DSP\ is due to isospin
breaking in the decay sequence of excited \D's, $f_{+-}\neq f_{00}$,
or $\tau_{\BM}<\tau_{\BZB}$ remains to be seen.

\begin{table}[ht]
\begintable
  $\B\sto$           &  ${\cal B}$ (\%)             \cr
  $\DZ X$            & $59.1\pm 2.3\pm2.1\pm1.6$    \nr
   $\DP X$           & $20.2\pm1.3\pm0.9\pm2.8$     \nr
   $\DSZ X$          & $25.1\pm1.9\pm1.2\pm1.7$     \nr
   $\DSP X$          & $20.6\pm1.5\pm0.9\pm0.7$     \cr
$\DZ_{\rm direct} X$ & $19.9\pm3.1\pm1.0$           \nr
$\DP_{\rm direct} X$ & $14.3\pm1.6\pm2.3$           \cr
   $\PS X$           & $1.10\pm0.05\pm0.08$         \nr
   $\PSP X$          & $0.28\pm0.05\pm0.05$ 
\endtable
\caption{Results on inclusive branching ratios.  The
second systematic error, when given, reflects the extrinsic
systematic from propagation of errors on branching ratios
used in reconstruction.}
\label{tab:inclus}
\end{table}

\vspace{2.3mm}
{\bf\boldmath \noindent \underline{$D\PI\PI$}}
\vspace{0.9mm}
\par
\begin{figure}[ht]
\hbox{\hfill\hskip0.1in\epsffile{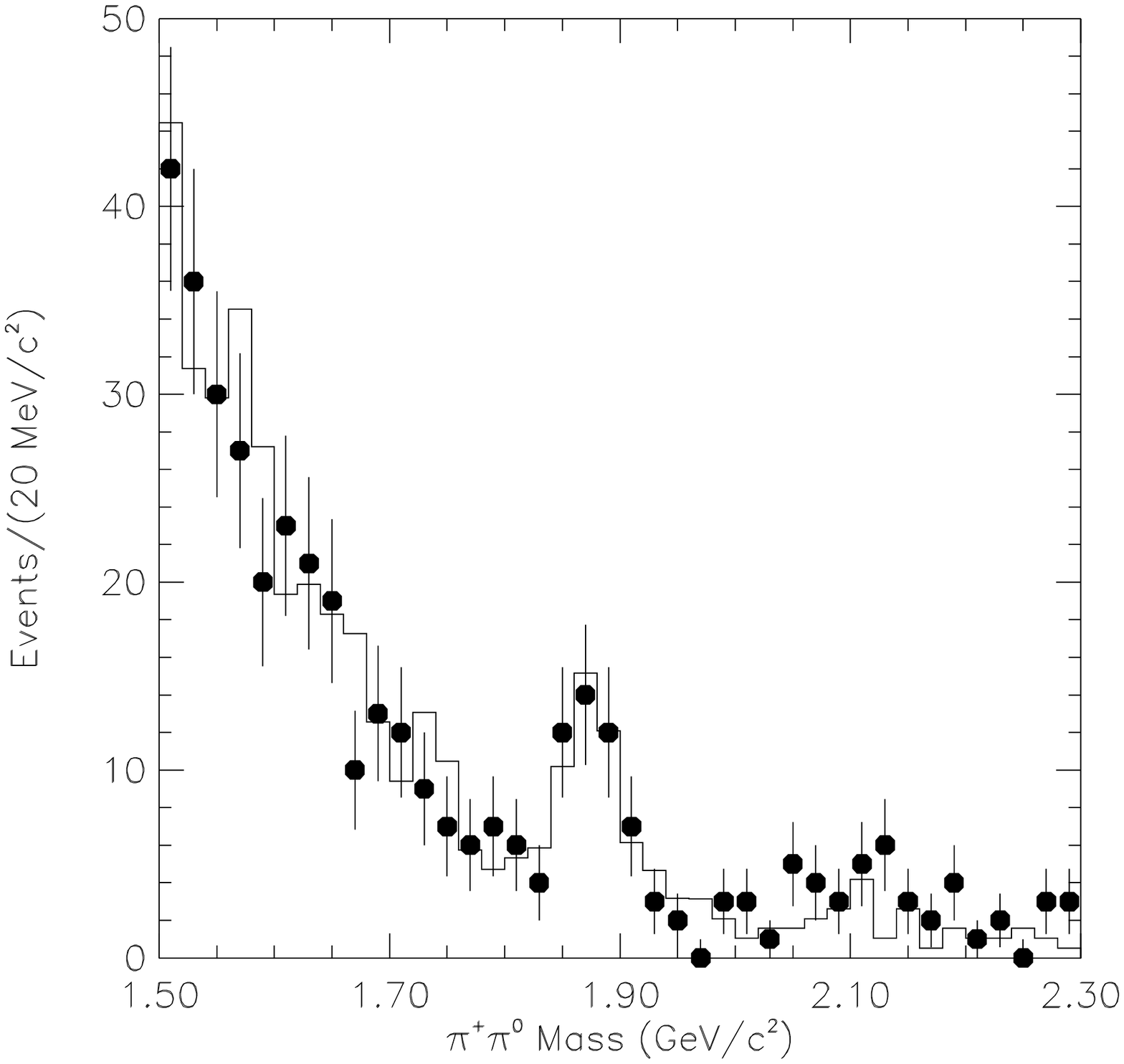}\hfill}
\caption{\PIZ\PIM mass; the peak at the \DP\ mass is
evident.  The background is from \K\RH\ and \KS\PI\@.
Data are the solid circles, the connected lines are Monte Carlo.}
\label{fig:ppp}
\end{figure}

\nobreak
The CsI calorimeter of CLEO-II has allowed the observation
of the decay modes $\DP\sto\PIP\PIZ$ and $\DZ\sto2\PIZ$.
The $\DP\sto\PIP\PIZ$ signal is shown in Fig.~\ref{fig:ppp}.
CLEO-II results on all three $\PI\PI$ decay modes are given
in Table~\ref{tab:ppp}. 

One can see from Table~\ref{tab:ppp} and the \D\ lifetimes
that $\Gamma(-+)/\Gamma(0+)=1.19\pm0.26$, which is rather
low for the \D\ system.  Before concluding that this
process is spectator driven, however, note
$\Gamma(00)/\Gamma(-+)=0.63\pm0.12$, which is similar
to the \K\ system, so an isospin analysis is appropriate.
The result of such an analysis is that the ratio of
$\Delta I=2$ to $\Delta I=0$ amplitudes, 
$|A_2/A_0|=0.72\pm0.13\pm0.11$, which is far greater
than the \K\ system, while 
$\delta_2-\delta_0=82^\circ\pm8^\circ\pm5^\circ$,
a large phase shift.

\epsfxsize=3.0in
\vspace{2.3mm}

\begin{table}[ht]
\begintable
  $\D\sto$           &  ${\cal B}$ (\%)             \cr
  $\PIM\PIP$         &  $0.136\pm0.012\pm0.012$     \nr
  $\PIZ\PIZ$         &  $0.086\pm0.016\pm0.015$     \nr
  $\PIZ\PIP$         &  $0.24\pm0.05\pm0.05$        \endtable
\caption{Results on $\D\sto2\PI$ branching ratios}
\label{tab:ppp}
\end{table}

\vspace{2.3mm}
{\bf \noindent\underline{\boldmath Wrong Sign Decays of the $D^0$}}
\vspace{0.9mm}
\par
\nobreak
We have observed a signal from tagged $\DZ$'s decaying
to $\KP\PIM$.  We tag the $\DZ$ with the charge of the
soft pion from $\DSP(\DSM)\sto\DZ\PIP(\DZB\PIM)$.  The
$\KP\PIM$ could either result from the doubly Cabibbo-suppressed
decay of the $\DZ$, or from $\DZ\sto\DZB$ mixing, followed
by Cabibbo-allowed decay of the $\DZB$.

The basic quantities of the analysis are the mass of the
putative $\KP\PIM$ system, $m_{\KP\PIM}$,
and the mass difference computed
by addition of the soft pion to this system, $\delta m$.
Backgrounds from $\K/\PI$ misidentification will tend
to peak in $\delta m$, but $m_{\KP\PIM}$ will not peak at $m_{\DZ}$; 
in fact, any $m_{\KP\PIM}$ that reconstructs near $m_{\DZ}$ under the
hypothesis that a misidentification occurred is cut.
Backgrounds from random slow pion tags
will tend not to peak in $\delta m$.

\begin{figure}[ht]
\hbox{\hfill\hskip0.1in\epsffile{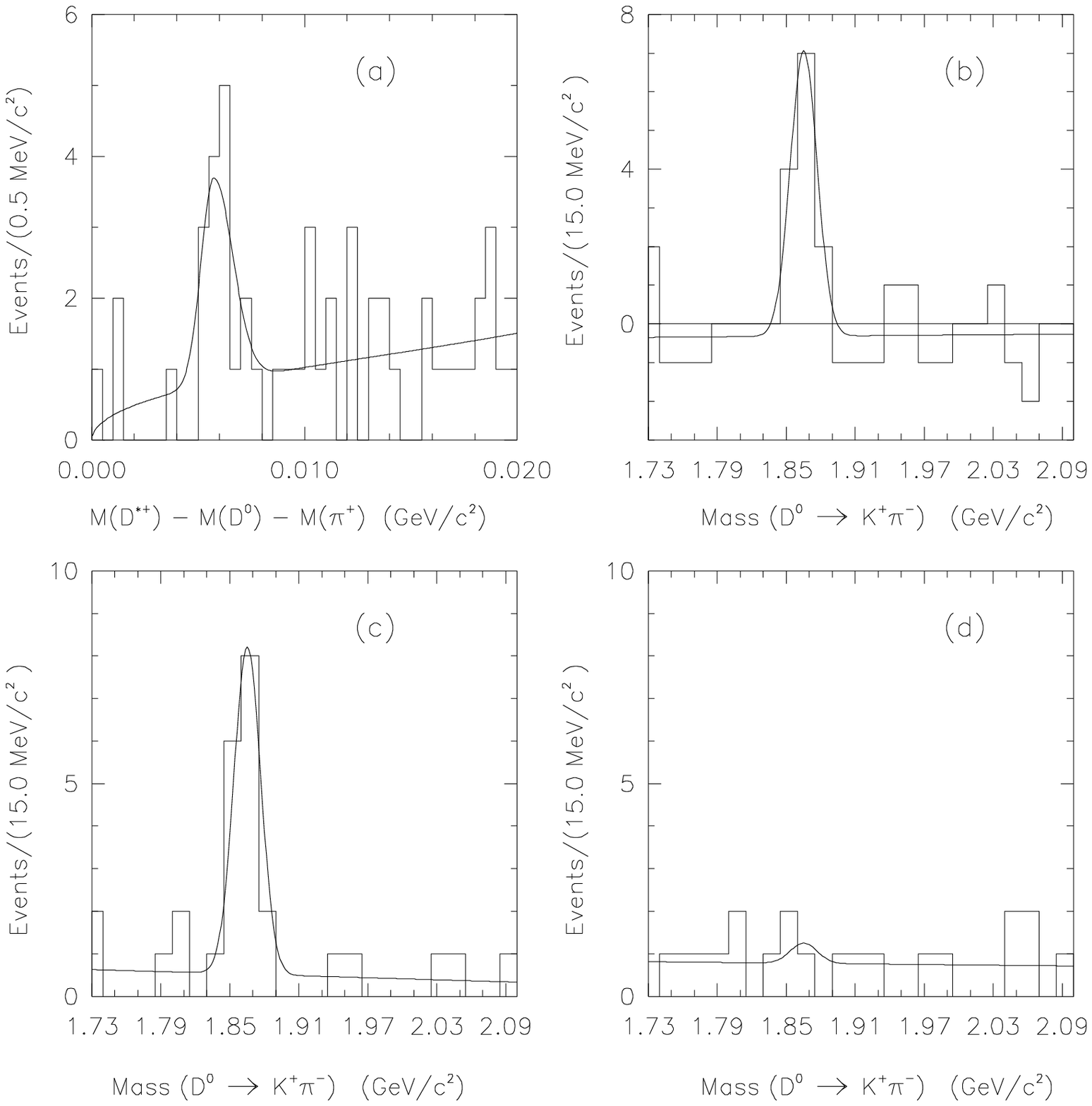}\hfill}
\caption{$\DZ\sto\KP\PIM$ candidates, (a) projected
on the $\delta m$ axis; (c) after a cut around the
expected $\delta m$, and projected on the $m_{\KP\PIM}$
axis, (d) taken from sidebands of $\delta m$ and projected
on the $m_{\KP\PIM}$ axis; and (b), the difference between
(c) and (d), showing the signal of 14.9 events.}
\label{fig:dcsd}
\end{figure}

The distribution of $\delta m$ for $\DZ\sto\KP\PIM$ candidates is
shown in Fig.~\ref{fig:dcsd}(a).  For this figure, hard $\K/\PI$
separation cuts have been made.  A signal region is defined
in $\delta m$, and the projection of this signal region on the
$m_{\KP\PIM}$ axis is shown in Fig.~\ref{fig:dcsd}(c).  Sidebands
in $\delta m$ are projected onto the $m_{\KP\PIM}$ axis
and shown in Fig.~\ref{fig:dcsd}(d); little peaking is evident.
The difference between (c) and (d) is the signal, shown
if Fig.~\ref{fig:dcsd}(b), and is 14.9 events on an expected
background of 0.9, a rather significant result.

\epsfxsize=3.0in
\vspace{2.3mm}

Having established the signal, the particle ID cuts are relaxed
to get a measure of the branching ratio.  The result is:
\begin{eqnarray}
R\equiv&{\Gamma({\DZ\{\sto \DZB\}\sto\KP\PIM})\over\Gamma({\DZ\sto\KM\PIP})}
= & \!\!\!\!\!\!\![0.77\pm0.25\pm0.25]\%\\
=& &\!\!\!\!\!\!\!\!\!\!\!\!\!\!\!\!\!\!\!(2.92\pm0.95\pm0.95)\tan^4{\theta_c}
\end{eqnarray}
where $\theta_c$ is the Cabibbo angle.

\vspace{2.3mm}
{\bf\boldmath\noindent\underline{Absolute 
$D^0$ and $D^+$ Branching Ratios}}
\vspace{0.9mm}

\nobreak
CLEO-II has also used the soft pions \DS\ decays to provide
the normalization in new, precise measurements of the
absolute branching ratios for $\DZ\sto\KM\PIP$ and
$\DP\sto\KM\PIP\PIP$.  The possibility of soft pion tags
from $\SIZB\sto\PIP\LCMB$ has been excluded in a Monte Carlo
independent way.  The result for the $\DZ$, where the
tags are only \PIP\ from the \DSP,
is:
\begin{equation}
{\cal B}(\DZ\sto\KM\PIP)=(3.912\pm0.082\pm0.17)\%
\end{equation}
The largest contribution to the systematic error results from
uncertainty in track reconstruction efficiency.

For the $\DP$ decay, the soft $\PIZ$ from the $\DSP$ must be
be used, which brings in background from $\DSZ\sto\DZ\PIZ$.
The analysis is specially designed to suppress systematic
uncertainty from the $\DS$ branching ratios.  The result is:
\begin{equation}
{\cal B}(\DP\sto\KM\PIP\PIP)=(10.0\pm0.5\pm0.7\pm1.4)\%
\end{equation}
The second systematic error results only from uncertainty
on the relative efficiency of soft $\PIZ$ to soft $\PIP$
reconstruction.

\vspace{2.3mm}
{\bf\noindent\underline{Conclusions}}
\vspace{0.9mm}
\par
\nobreak
CLEO-II's large data sample has been exploited to further
understanding of a number of hadronic weak decays of heavy
mesons.  There is clear evidence that a number of two
body branching ratios for the \BM\ are larger than the
analogous branching ratios for the \BZB\@.  It remains
to be seen whether $\tau_{\BM}<\tau_{\BZB}$.

\vspace{2.3mm}
{\bf \noindent\underline{Acknowledgements}}
\vspace{0.9mm}

\nobreak
Many thanks to John Carr for his tolerance and persistence.

\vspace{2.3mm}
{\bf \noindent\underline{References}}
\vspace{0.9mm}
\par
\def\PRref#1&#2&(#3)#4{\unskip\ #1~\bf#2\rm, (#3) #4}
\def\ZPC{\it Z. Phys. C}
\def\etal{{\it et al.}}
\noindent
[1] M. Bauer, B. Stech, and M. Wirbel, \PRref\ZPC&29&(1985)637; 
{\it ibid} {\bf 34}, (1987) 103; {\it ibid} {\bf 42}, (1989) 671.

\end{document}